\documentclass[twocolumn,trackchanges]{aastex701}

\usepackage{amsmath}
\usepackage{soul}
\usepackage{ulem}

\begin{document}

\title{The dependence of Circumgalactic Medium properties on halo assembly histories in the IllustrisTNG simulations}

\author[orcid=0009-0001-0716-9867]{Yi-yuan Zhang}
\affiliation{Institute for Astronomy, the School of Physics, Zhejiang University, Hangzhou 310027, People's Republic of China} 
\email{12445050@zju.edu.cn}

\author[orcid=0000-0001-8426-9493]{Hou-Zun Chen $^\dagger$} 
\affiliation{Institute for Astronomy, the School of Physics, Zhejiang University, Hangzhou 310027, People's Republic of China}
\email[show]{chenhz\underline{ }zju@zju.edu.cn}

\author[orcid=0000-0002-5458-4254]{Xi Kang $^*$}
\affiliation{Institute for Astronomy, the School of Physics, Zhejiang University, Hangzhou 310027, People's Republic of China}
\affiliation{Purple Mountain Observatory, 10 Yuan Hua Road, Nanjing 210034, People's Republic of China}
\email[show]{kangxi@zju.edu.cn}

\begin{abstract}

While halo mass is the dominant factor shaping the embedded galaxies, the properties of the circumgalactic medium (CGM) also depend on halo assembly history. To investigate this, we calculate the formation times for TNG50 halos with masses between $10^{10.5}$ and $10^{12.5} M_\odot$, classifying them into `early-' and `late-forming' populations. It is found that across all mass bins, early-formed halos generally host galaxies with higher stellar mass and higher metallicity, with lower CGM gas mass and lower specific star formation rate (sSFR) at $z\sim0$. 
For the CGM metallicity, `early' halos with masses below $10^{12}\mathrm{M_\odot}$ show systematically higher gas phase metallicities, whereas in the $10^{12-12.5}\mathrm{M_\odot}$ bin the trend reverses. When examining the origins of the CGM gas, it is found that fresh accretion is insensitive to assembly history, whereas the `late' galaxies experience more wet mergers. These differences in gas properties arise from processes after the formation time, given that the CGM gas masses show no significant differences at formation time. Finally, our analysis of CGM kinematics shows that for halos below $10^{12}\mathrm{M_\odot}$, the cold gas in late-forming halos carries higher specific angular momentum and simply has a higher degree of rotational support, while the same properties in the $10^{12-12.5}\mathrm{M_\odot}$ bin shows no significant dependence on assembly history.

\end{abstract}


\section{Introduction}
\label{sec1:introduction}

As an extended gasous halo surrounding galaxy, the circumgalactic medium (CGM) bridges the interstellar medium (ISM) and intergalactic medium (IGM), acting as a crucial reservoir that regulates galaxy formation and evolution \citep[see][for a comprehensive review]{sec1araa2017,sec1araa2023}. On the one hand, fresh accretion from the IGM passes through the CGM before reaching the galaxy, providing the fuel required for star formation. On the other hand, feedback from supernova and AGN can inject energy into the CGM gas and reshape the distribution of baryon or even dark matter within the halo. Owing to the vital role of the CGM, extensive observational efforts have been done over the past decades, for example large optical spectroscopic surveys with robustly calibrated galaxy physical properties such as the Sloan Digital Sky Survey \citep[SDSS;][]{sec1sdss}, UV all-sky imaging surveys such as Galaxy Evolution Explorer \citep[GALEX;][]{sec1galex}, moderate resolution but with improved sensitivity spectrographs such as Cosmic Origins Spectrograph \citep[COS;][]{sec1cos}, observatories which operate in the X-ray band and are capable of probing the diffuse hot gas in halos such as Chandra X-ray observatory \citep[CXO][]{sec1cxo} , XMM-Newton \citep[][]{sec1xmmnewton}, and eROSITA X-ray telescope \citep[][]{eROSITA2021A&A...647A...1P}. By measuring emission lines from CGM itself and absorption features imprinted on background AGN/QSO sigitlines \citep[][]{sec1araa2017}, these observational programs have accumulated staticstically meaningful samples and provided powerful constraints on the thermodynamic, chemical, and kinematic properties of the CGM.\citep[][]{sec1meaningfulsampleCookesy2010,sec1meaningfulsampleProchaska2011,sec1meaningfulsampleTumlinson2013,sec1meaningfulsampleLehner2014,sec1meaningfulsampleLiang2014,sec1meaningfulsampleWerk2016,sec1meaningfulsampleLehner2020,sec1meaningfulsampleLi2020,sec1meaningfulsampleZhang2025}.

Interpreting observables, such as luminosities or absorption-line depths, often requires modeling assumptions to convert them into physical quantities. Moreover, observations can hardly trace the evolutionary path of galaxies as they are limited to a single point in time. Motivated by these limitations, a variety of hydro-simulation approaches have been developed to complement the observations. Cosmological hydrodynamic simulations - such as IllustrisTNG \citep[][]{TNG100,TNG50Nelson,TNG50Phillepich}, EAGLE \citep[][]{EAGLE}, and the forthcoming COLIBRE \citep[][]{COLIBRE1,COLIBRE2} - provide self-consistent cosmological environments for studying galaxy populations across large, statistically representative samples. Zoom-in simulations, such as FIRE \citep[][]{FIRE}, Auriga \citep[][]{Auriga}, and NIHAO \citep[][]{NIHAO}, achieve much higher resolution by sacrificing coverage of large scale environments, enabling more detailed studies of individual galaxies. In addition, idealized hydrodynamic simulations also contribute important insight into the smaller-scale CGM processes such as the formation and destruction of cold cloud \citep[][]{sec1idealsimBirnboim2003,sec1idealsimFielding2017,sec1idealsimGronke2018,sec1idealsimStern2019,sec1idealsimStern2020,Barbani2023MNRAS.524.4091B}. Regardless of their specific methodology, these simulations have successfully reproduced many observed CGM phenomena and scaling relations within their respective regimes, such as the HI column densities and covering fractions, cold gas mass fractions \citep[][]{sec1reproduceobsGutcke2017,sec1reproduceobsHafen2017,sec1reproduceobsNelson2018,Hafen2019,traceruse2020}, and facilitated the study of the redshift evolution of these systems. Nevertheless, simulations usually underpredict OVI column densities in dwarfs and MW-like galaxies \citep[][]{Hummels2013MNRAS.430.1548H,Gutcke2017MNRAS.464.2796G}, although some AGN models may be able to mitigate this \citep[][]{Oppenheimer2018MNRAS.474.4740O}.

The primary factor governing the properties of the CGM is the halo mass, or equivalently, the gravitational potential. The potential well would shape the temperature and pressure structure of the CGM \citep[][]{sec1idealsimFielding2017,sec1haloshapeLochhaas2020}, and regulates the gas accretion mode of halo \citep[classically,][]{Dekel2006}. In addition, because the escape velocity scales with the circular velocity, which itself scales approximately as $M_{\rm vir}^{1/3}$, the gravitational potential determines the fate of winds and mass flows: whether they escape the halo entirely or re-accreted later as the form of cycling or galactic fountain \citep[][]{sec1haloshapeOppenheimer2008,sec1haloshapeOppenheimer2010,sec1haloshapeMuratov2015,AA2017}. Works such as \cite{sec1haloshapemetalTumlinson2011} and \cite{sec1haloshapemetalProchaska2017} have demonstrated that the distribution of metals in the CGM is strongly correlated with halo mass.

If this gas reservoir feeding galaxy formation were uniquely determined by the halo mass, galaxy properties would be also uniquely determined by the halo mass. In fact, the star-forming and passive galaxies exhibit distinct CGM ionization states even within narrow mass ranges \citep[][]{sec1haloshapemetalTumlinson2011,Bordoloi2018ApJ...864..132B}, together with the scatter of the stellar mass–halo mass (SMHM) relation, indicates that additional factors must be taken into account when investigating the CGM properties. Among these probabilities, AGN feedback from supermassive black holes (SMBHs), is one of the most widely discussed factors \citep[][]{sec1smbhNelson2015,sec1smbhOppenheimer2018,sec1smbhDavies2020,sec1smbhTruong2020,sec1smbhZinger2020}. On the one hand, AGN can inject large amounts of thermal energy, heating gas over extended regions, on the other hand, it can efficiently drive material into or even out of the CGM, thereby enriching the circumgalactic environment with metals from inside the galaxy. Furthermore, the large scale environment in which a galaxy resides can also influence the CGM gas properties such as the accretion mode \citep[][]{sec1envshapeKeres2005,Dekel2006,AragonCalvo2019OJAp....2E...7A,Gal2023A&A...671A.160G} or the angular momentum \citep[][]{Wang2022MNRAS.509.3148W,Lu2022MNRAS.509.2707L}.

While examining the origin of the scatter in the relationship between the gas fraction and halo mass for present-day $L_*$ central galaxies using the EAGLE simulation, \cite{D19} found that at fixed halo mass, the CGM gas fraction $f_{\mathrm{CGM}}$ correlates strongly with black hole mass and intrinsic binding energy of the inner halo, which serves as a proxy for halo assembly time. Specifically, halos that assemble earlier exhibit systematically lower present-day CGM mass fractions than their later-assembling counterparts. They attribute this difference primarily variations in the cumulative feedback energy. 
To control the environmental effects and effectively isolate the impact of assembly history on the internal evolution, and consequently on CGM properties, \cite{D21} (hereafter D21) selected a halo sample with $M_{h} \approx 10^{12} \mathrm{M_\odot}$ at $z=0$ and performed genetically modified (GM) simulations to investigate how the assembly history affects galaxy properties. They showed that artificially shifting the assembly history to earlier times leads to a pronounced suppression of the specific star formation rate (sSFR), a reduction in both the mass and density of the CGM, and a more massive central black hole.

While this study clearly demonstrates how assembly history influences galaxy properties from an in-situ perspective, it is essential to recognize that, within a cosmological framework, this history is also intrinsically linked to the large-scale structure (LSS) environment.
As \cite{Gao2005} has demonstrated, halos that assemble earlier are more clustered (i.e. so-called 'assembly bias'). This implies that early-forming halos tend to reside near more massive neighboring halos and the accretion of gas from the IGM could be affected by nearby environment. Consequently, their CGM and galaxy properties may differ systematically from those of late-forming halos not only by internal processes, but also due to differences in their environmental gas supply.

Motivated by these considerations, we adopt TNG50, the highest resolution simulation in the IllustrisTNG suite, to obtain a statistically robust sample in a self-consistent cosmological context while retaining a resolution high enough to resolve the bulk properties of CGM in most galaxies. With this setup, we select a sample of 4256 halos with masses between $10^{10.5-12.5} \mathrm{M_\odot}$, and we divide them into four mass bins (with 0.5 dex in width) to systematically assess how assembly history affects CGM properties across different mass scales.

This paper is organized as follows. In section~\ref{sec2:Methods}, we introduced the TNG50 simulation and the Monte Carlo tracer particles used within the moving-mesh framework. Then we describe our sample selection and classification, as well as the definition of frequently used key physical quantities. In section~\ref{sec3:result} we present our main results and then provide a preliminary explanation for these differences in section~\ref{sec4:discussion}. Finally, we give a summary in section~\ref{sec5:summary}.

\section{Methods}
\label{sec2:Methods}

\subsection{The TNG50 simulation}
\label{subsec2.1:tng50}
TNG50 \citep[][]{TNG50Nelson,TNG50Phillepich} is the third and final simulation of the IllustrisTNG suite \footnote{\url{https://www.tng-project.org/}} \citep[hereafter TNG;][]{TNG100,TNG1,TNG2,TNG3,TNG4,TNG5,TNGdata}, which is an ongoing series of large, cosmological, magnetohydrodynamic simulation performed with the moving-mesh code \texttt{Arepo} \citep[][]{arepo2010} and including a comprehensive set of physical models for the formation and evolution of galaxies across cosmic time. All TNG runs contain 100 snapshots, which start from cosmologically motivated initial conditions assuming an updated cosmology consistent with the Planck 2015 results ($\Omega_{\Lambda,0}=0.6911, \Omega_{m,0}=0.3089, \Omega_{b,0}=0.0486, \sigma_8=0.8159, n_s=0.9667, h=0.6774$) and are simulated from $z = 127$ to $z = 0$, respectively. The results of the simulation have been verified in agreement with many observed properties and scaling relations\citep[][]{TNG1,TNG2,TNG3,TNG4,TNG5}.

In TNG project, there are three distinct simulation volumes: TNG50, TNG100 and TNG300, which employs a periodic box of side length $\sim 50$, 100, 300 Mpc and includes $2 \times 2160^3$, $2 \times 1820^3$, $2 \times 2500^3$ resolution elements respectively. Among them, TNG50-1 offers the highest mass and spatial resolution, with an average mass resolution of $8 \times 10^4 \mathrm{M_\odot}$ for baryon and $4.5 \times 10^5 \mathrm{M_\odot}$ for dark matter, respectively. The high resolution of TNG50 enables the analysis over a large sample size of galaxies while retaining a fully cosmological and well-resolved environment context. Moreover, since the diversity of assembly history are most pronounced in halos below $10^{12.5}\mathrm{M_{\odot}}$, the sample size in TNG50-1 simulation is large enough for the analyze in our work (refer to Section \ref{subsec2.3:samplec_selection}).

Although TNG50 has achieved high resolution, processes operating on smaller scales - such as star and AGN formation and feedback, chemical enrichment, radiative effects and so on - still require the use of so-called subgrid models to be properly resovled. The detailed descriptions of these subgrid models can be found in a set of papers \citep[][]{subgridmodel,subgridmodelbh}, and here we provide a brief overview of the aspects most relevant to our analysis. (i) Star formation is modeled with a density-dependent prescription. The gas whose density exceeds a threshold (i.e. star-forming) can stochastically form stars with a probability; (ii) Stellar population evolution and chemical enrichment following supernovae Ia, II, as well as AGB stars, with individual accounting for the nine elements H, He, C, N, O, Ne, Mg, Si, and Fe; (iii) Supernovae feedback can release its energy in the form of a kinetic, galactic-scale wind. (iv) The ISM model employs a pressurization model for star-forming gas, which imposes a temperature floor of $\sim 10^4 \mathrm{K}$. Therefore, the temperature of ISM gas is an effective temperature rather than a physical temperature. (v) A black hole (BH) particle with a mass of $8 \times 10^5 h^{-1} \mathrm{M_\odot}$ is seeded in the center of a dark matter halo that is both more massive than $5 \times 10^{10} h^{-1} \mathrm{M_\odot}$ and devoid of any pre-existing BH particles. Then, the BH would grow with an eddington-limited Bondi accretion. (vi) The BH operates in two distinct feedback modes determined by its accretion rate. When the accretion ratio $\dot{M}_{\rm BH} /\dot{M}_{\rm Edd}$ exceeds the threshold $\chi$, the BH releases thermal energy into the surrounding gas cells (referred to as ``thermal mode"). When the accretion ratio falls below $\chi$, the AGN instead injects feedback energy as a momentum kick in a randomly chosen direction (viz. ``kinetic mode"). The ratio $\chi$ is defined as: 
\begin{equation}
    \chi =\min\left[0.002\left(\frac{M_{\rm BH}}{10^8 M_{\odot}}\right)^2,0.1\right]
\end{equation}

\subsection{Monte Carlo Tracers}
\label{subsec2.2:tracer}
The TNG simulations employ a moving-mesh approach. In order to trace the realistic evolution of gas over time, a ``Monte Carlo" (MC) tracer particle technique is implemented \citep[][]{tracermethod2013,tracermethod2013Nelson,traceruse2020}. Briefly, all cells are initially populated with an equal number of tracers, and each tracer is subsequently transferred between cells with a probability determined by the local mass flux. In this way, this tracer method self-consistently follows the transfer of mass among various baryon components. In particular, tracer particles can transfer:\\
(i) from gas cell to gas cell via finite volume fluxes, refinement or derefinement.\\
(ii) from gas cell to star particle via star formation. \\
(iii) from gas cell to wind phase particle via galactic winds driven by the stellar feedback.\\
(iv) from gas cell to BH particle via the BH accretion.\\
(v) from star particle or wind phase particle to gas cell via the stellar mass return or recoupling stellar wind.\\
(vi) from BH particle to BH particle via the BH mergers .\\
Finally, we should note that MC tracers sometimes can also be numerically diffusive \citep{tracermethod2013}.

\subsection{Sample Selection}
\label{subsec2.3:samplec_selection}
Given the mass resolution level of TNG50-1 simulation, the results when analyzing systems with mass lower than $10^{7.5} \mathrm{M_\odot}$ may be affected by the resolution effects \citep[][]{TNG50Phillepich}. In addition, the sample size of central halos with masses above $10^{12.5}\mathrm{M_\odot}$ is limited by the relatively small simulation box of TNG50-1, making it difficult to perform a statistically robust analysis. In the framework of hierarchical structure formation, such massive halos typically form relatively late, resulting in only modest differences in their formation times and weak assembly bias \citep[][]{Gao2005,Sunayama2016MNRAS.458.1510S}.
Therefore, taking into account resolution considerations, sample size, and the aspiration to cover the widest possible mass range, we select data from the TNG50-1 simulation based on the following criteria:\\
(i) The galaxy is a central galaxy of a halo.\\
(ii) The mass of the halo is within the range of  $10^{10.5}-10^{12.5} \mathrm{M_\odot}$.\\
(iii) The total gas mass - without restricting it to CGM gas alone - exceeds $10^7 \mathrm{M_\odot}$, corresponding to roughly 100 gas cells to avoid resolution effects \citep[][]{TNG100,TNG50Nelson,shotnoise2020}.

These criteria yield a sample of 4256 galaxies in total, and we divide them into four bins of 0.5 dex, which containing 2622, 1104, 381 and 149 samples respectively.

\subsection{Physical definition}
\label{subsec2.4:physical_definition}
The CGM is commonly defined as the gas residing between the ISM and IGM. It is often delineated as the region extending from 0.1-0.15 $R_{\rm vir}$ out to $R_{\rm vir}$, with $R_{\rm vir}$ typically taken to be $R_{\rm 200c}$ in simulations. It is important to emphasize that this boundary is introduced purely for practical purposes and does not imply any sharp physical transition in the gas properties.

In this work, we adopt $[0.15,\ 1.0]R_{200c}$ as the extent of the CGM, where $R_{200c}$ is the radius of a sphere centered at the center of the halo whose mean density is 200 times the critical density of the universe. This choice greatly minimizes contamination from the ISM gas, which contains a large fraction of star forming gas whose temperature, as discussed in Section~\ref{subsec2.1:tng50}, is a ``effectiive" temperature of the equation of state rather than a physical one.
In addition, we introduce the following classifications:
\begin{itemize}
    \item Formation time: the formation time of a halo is defined as the cosmic time or redshift at which the halo attains half of its present-day ($z=0$) mass.\\ 
    \item Classification between `early' and `late': sorted by the formation time of each halo, we classify the first 33\% (earliest-forming) and the last 33\% (latest-forming) of halos as ‘early’ and ‘late’, respectively. It is important to note that massive halos typically assemble later than low-mass halos in hierarchical structure formation, our classification of `early' and `late' halos are performed within each mass bin.\\

    \item Classification between different gas phases: CGM gas is classified into cold phase ($T < 10^{4.5}\rm K$), warm phase ($10^{4.5}{\rm K} < T < 10^{5.5}\rm K$), and hot phases ($T > 10^{5.5}\rm K$) according to its temperature. This classification is also similar to that used by previous CGM studies based on the TNG simulations \citep[][]{Ramesh2023MNRAS.518.5754R,Ramesh2023MNRAS.522.1535R}.\\
\end{itemize}

\begin{figure}
    \centering
    \includegraphics[width=\linewidth]{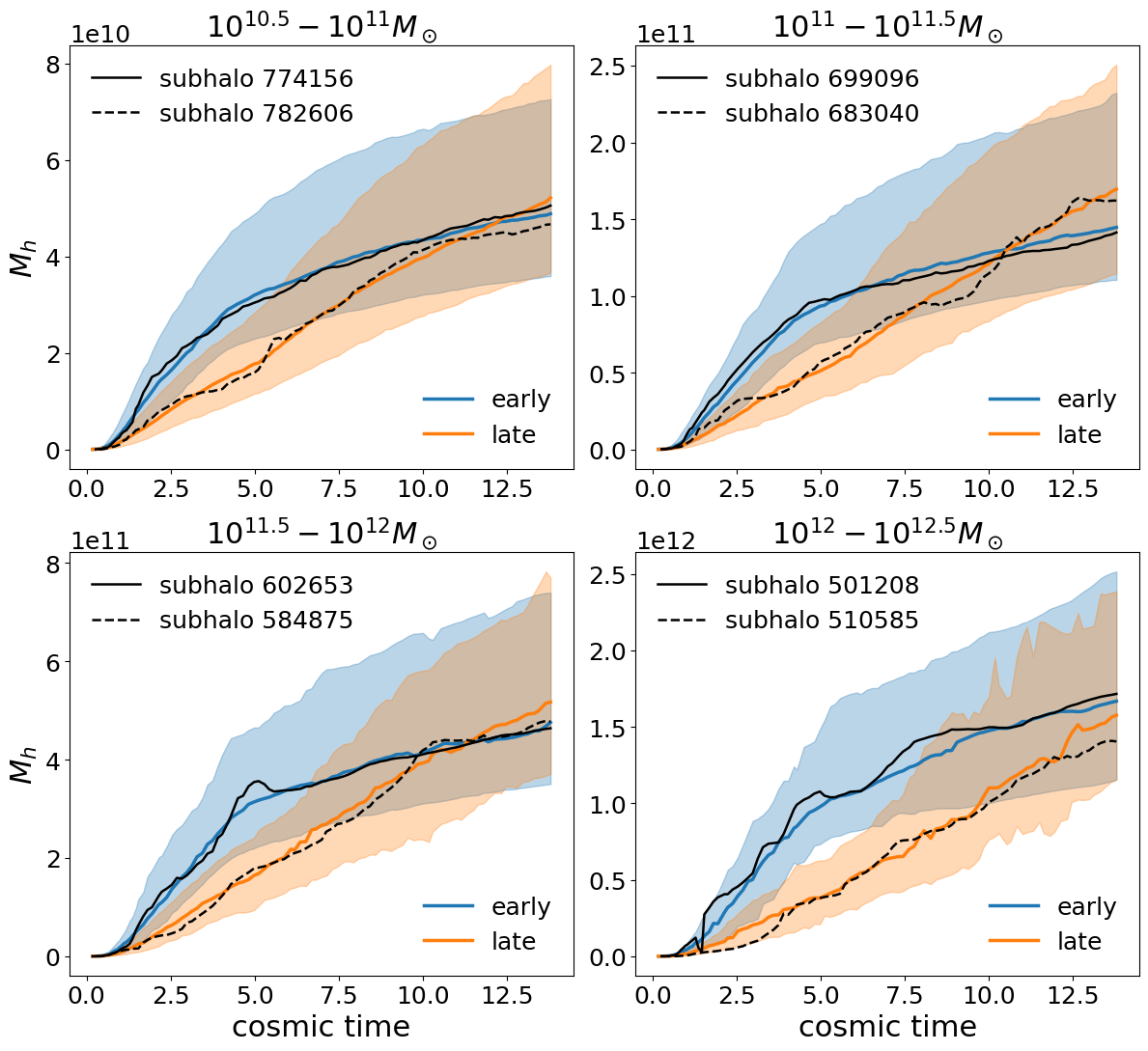}
    \caption{The median mass accretion histories for `early' and `late' halos. In each panel, the blue curve corresponds to halos with earlier assembly histories (`early’), while orange curve represents those with later assembly histories (`late’). Shaded regions indicate the 16-84th percentile range. Moreover, solid and dashed black lines show the specific accretion history for individual `early' and `late' halos, respectively}
    \label{fig:accretionhistory}
\end{figure}

To demonstrate the differences in assembly history more explicitly, we show in Fig.~\ref{fig:accretionhistory} the median mass accretion histories for `early' and `late' halos in each mass bin. As expected, `early' halos tend to assemble a larger fraction of their mass at earlier times, while `late' halos exhibit more extended mass growth toward later times.

\section{Result}
\label{sec3:result}

\begin{figure*}[!htbp]
\gridline{
    \includegraphics[width=0.33\textwidth]{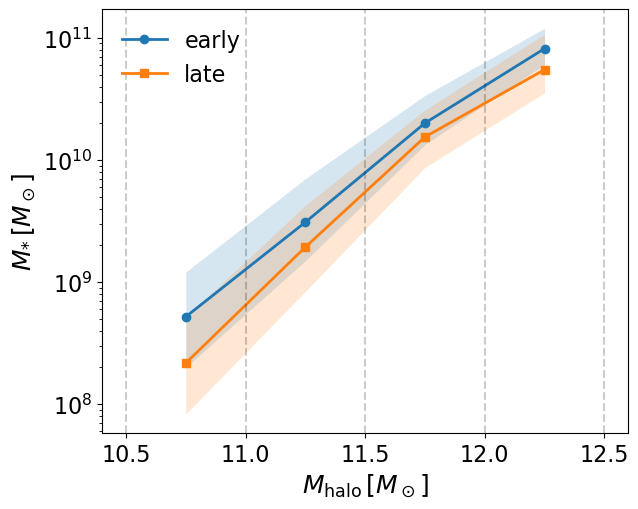}
    \includegraphics[width=0.33\textwidth]{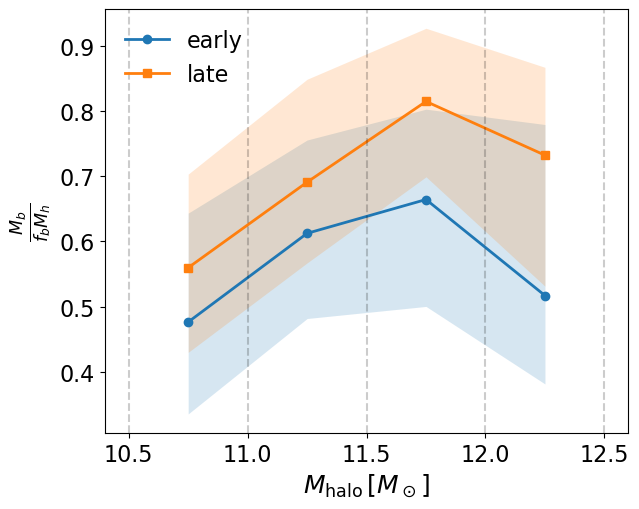}
    \includegraphics[width=0.33\textwidth]{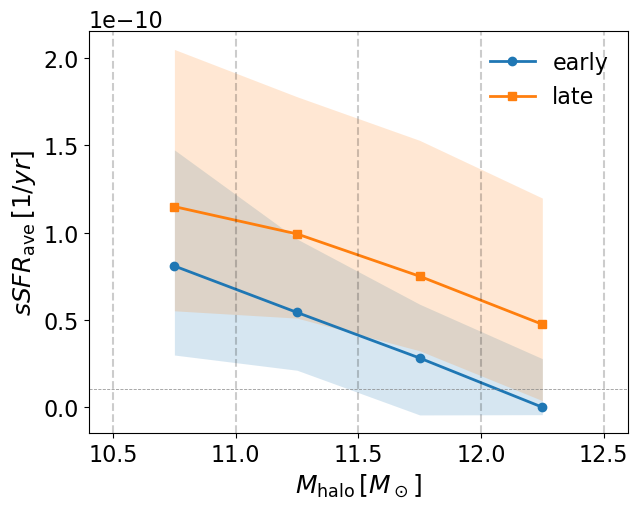}
}
\caption{The global baryonic properties of the halo. From left to right, the panels show the total stellar mass, the baryon fraction, and the specific star formation rate at $z=0$, respectively. In each panel, the blue symbols mark the median values of halos with earlier assembly histories (‘early’), while orange symbols represent those with later assembly histories (‘late’). The shaded regions show the 16-84th percentile range of the sample distribution.}
\label{fig:halo_property}
\end{figure*}

\subsection{The difference of global baryonic properties}
\label{subsec3.1:global_properties_of_the_halo}

We begin by examining whether halos with different assembly histories exhibit different global baryonic properties. 
Fig.~\ref{fig:halo_property} shows three global baryonic properties between `early' and `late' halos across the four mass bins. From left to right, the panels show the total stellar mass, the baryon fraction, and the specific star formation rate at $z=0$, respectively. Specially, the sSFR is derived from the differential of stellar mass between two adjacent snapshots rather than using the instantaneous SFR from the gas cells directly. 

From Fig.~\ref{fig:halo_property}, it is found that early-formed halos tend to have slightly larger stellar mass, lower baryon fraction and they have converted a larger fraction of their baryon budget into stars (i.e. $M_* / M_b$, not shown here for clearity). This trend of higher stellar mass is consistent with the results reported by \cite{Tojeiro2017MNRAS.470.3720T,Zehavi2018ApJ...853...84Z} and \cite{Montero2021MNRAS.508..940M}, who used the semi-analytic model L-GALAXIES and two samples ($M_h=10^{11.5}M_\odot$ and $10^{12.5}M_\odot$) drawn from TNG300, respectively. However, their present-day specific star-formation rates are systematically lower; for halos in the $10^{12-12.5} \mathrm{M_\odot}$ bin, about 70\% of the host galaxies exhibit an sSFR $<10^{-11}\rm yr^{-1}$, indicating that they are essentially quenched. This suppression of sSFR is consistent with \cite{D21}. More noticeable, by defining an observational proxy for the formation time, \cite{Lim2016MNRAS.455..499L} found that for central galaxies with stellar masses below $10^{11}M_\odot$, earlier formation times are associated with higher quenching fractions.
This is naturally explained by their earlier star formation histories: ``early" halos have formed stars over a longer period of time and have therefore experienced more extended episodes of stellar and AGN feedback, which suppress their recent star formation rate and, in the most massive cases, drive them into quiescence.

\begin{figure}
\gridline{
    \includegraphics[width=0.45\linewidth]{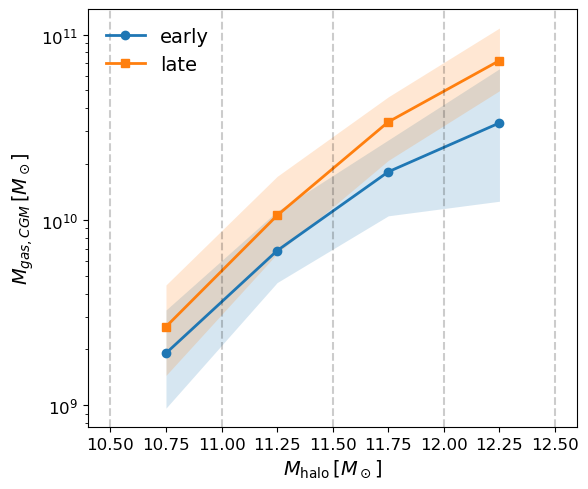}
    \includegraphics[width=0.45\linewidth]{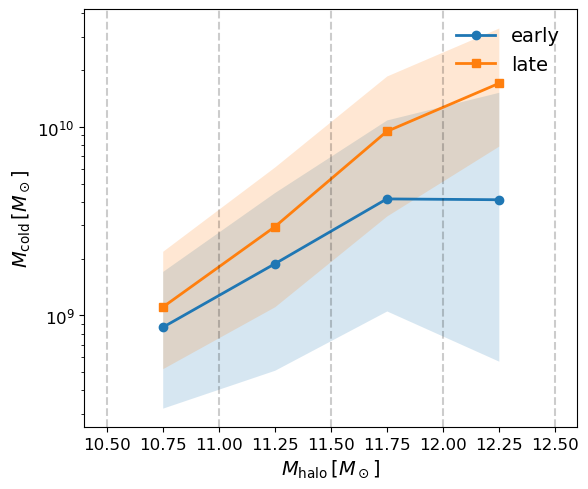}
}
\gridline{
    \includegraphics[width=0.45\linewidth]{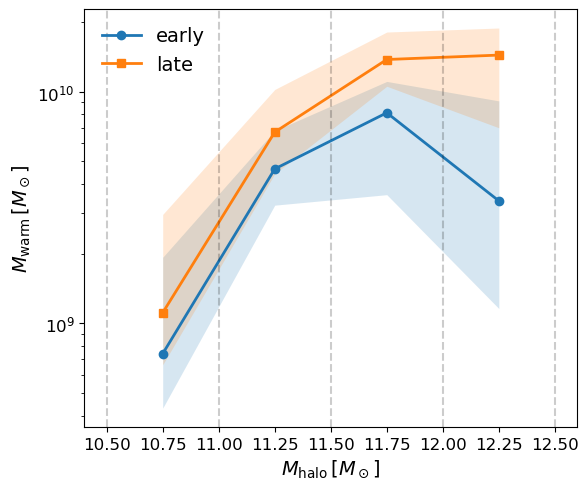}
    \includegraphics[width=0.45\linewidth]{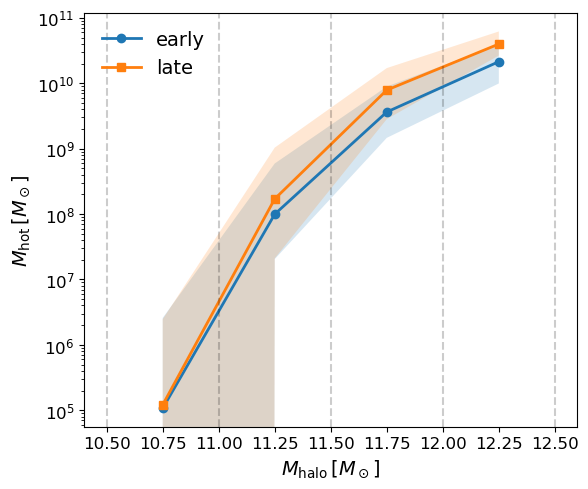}
}
\caption{Total and phase-resolved mass of CGM gas. Clockwise from the top-left, the panels show the total CGM gas mass, and the masses of the cold, hot, and warm phases. Symbols follow the same conventions as in Fig.~\ref{fig:halo_property}.}
\label{fig:cgm_thermo_property}
\end{figure}

\subsection{Thermodynamic and mass compositon of the CGM}
\label{subsec3.2:thermal_properties_of_the_cgm_gas}

Fig.~\ref{fig:cgm_thermo_property} presents the total CGM gas mass, as well as the mass budget in cold, warm and hot phases. In all cases - both the total mass and each phases - the `late' halos exhibit slightly larger CGM gas mass than `early' halos. This trend may arise for two reasons, which will be discussed in detail in Section~\ref{sec4:discussion} . Firstly, `early' halos have experienced more extended periods of star formation, which have consumed a larger fraction of their gas reservoir. Secondly, `late' halos can acquire additional gas after the formation time through intergalactic transfer and mergers, whereas `early' halos can only gain predominantly through gradual fresh accretion.

It is seen from the plot that at the present epoch, `late' halos retain more CGM gas, particularly more cold and warm gas. This provides a more abundant fuel supply for ongoing star formation in their central galaxies, resulting in higher sSFR at $z=0$ (shown in the lower right panel of Fig.~\ref{fig:halo_property}), even though their total stellar mass has not yet caught up with those of `early' halos.

\begin{figure*}
    \includegraphics[width=0.95\textwidth]{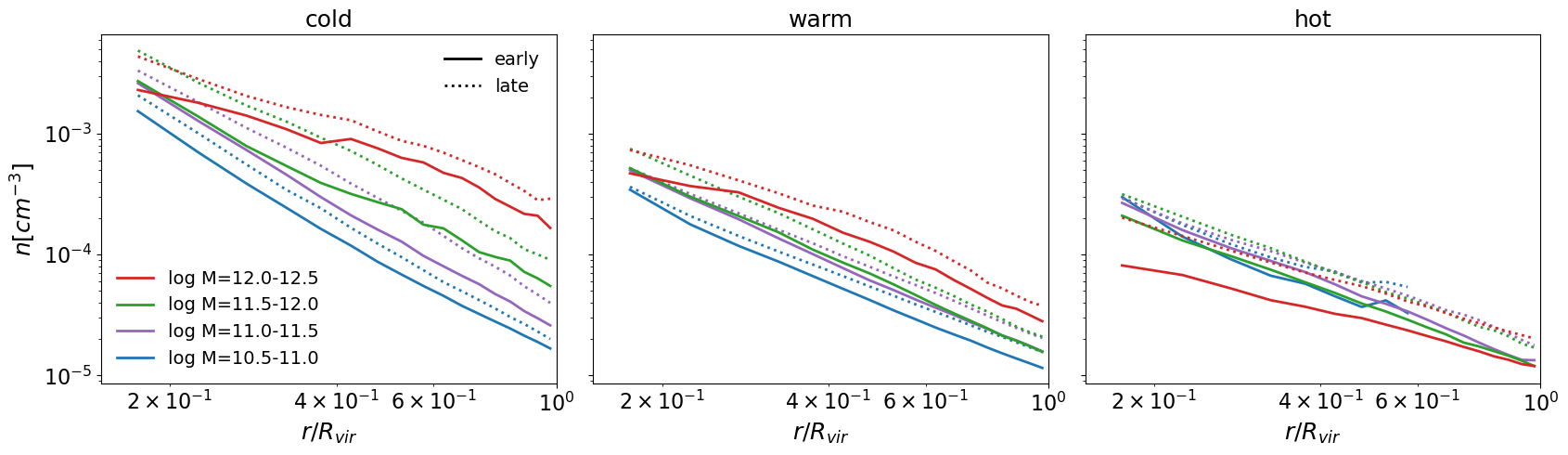}
    \caption{The radial profiles of the density for the different gas phases, the curves represent the median value. From left to right, the panels show the cold, warm, and hot phases. Different colors indicate different halo masses and different line styles represent distinct formation time, specifically, solid lines correspond to early-forming halos, while dotted lines represent late-forming ones.}
    \label{fig:rhoprofile}
\end{figure*}

The density profiles of CGM gas for `early' and `late' halos are examined in Fig.~\ref{fig:rhoprofile}. The density slopes for `early' and `late' halos are nearly identical. This indicates that the impact of assembly history on the CGM gas is primarily an overall mass offset rather than a structural change in the spatial distribution of the gas. Note that for halos in the $10^{10.5-11}\mathrm{M_\odot}$ mass bin, the hot phase is essentially absent in the outer CGM. This is a consequence of our temperature-based phase definition, in which hot gas is defined as having $T > 10^{5.5}\rm K$. This temperature is comparable to the virial temperature of halos in this mass bin, implying that their gravitational potential is insufficient retain such hot gas in the outer regions of the CGM.

\subsection{Metal enrichment and distribution in the CGM}
\label{subsec3.3:metal}

The origins of gas in the CGM are diverse. On the one hand, CGM gas could be from the outflow driven by supernovae or AGN feedback in the galaxy, on the other hand, it could be supplied through continuous accretion from IGM in the cosmic web. Because of the intertwined origins, it is generally difficult to disentangle inflows from outflows based solely on observable gas properties such as temperature or density \citep[][]{sec1haloshapeOppenheimer2010,sec33gasonlycannotLehner2013,sec33gasonlycannotWerk2014,sec33gasonlycannotFord2016,sec1araa2017,Hafen2019}.

However, metals are predominately produced inside galaxies through stellar nucleosynthesis and are transported into the CGM via stellar winds, supernova explosions, and other feedback processes. Metallicity therefore serves as a natural indicators of galaxy-processed material and recycled gas \citep[][]{sec1haloshapeOppenheimer2008,sec33metaltracePeeples2014,Hafen2019}.

\begin{figure*}
\gridline{
    \includegraphics[width=0.45\textwidth]{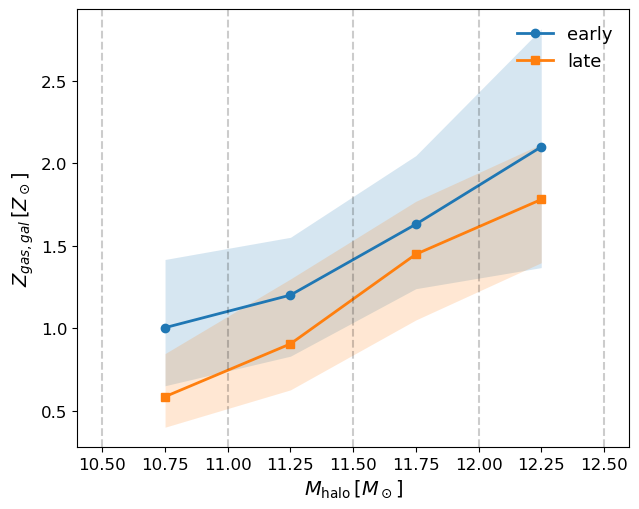}
    \includegraphics[width=0.45\textwidth]{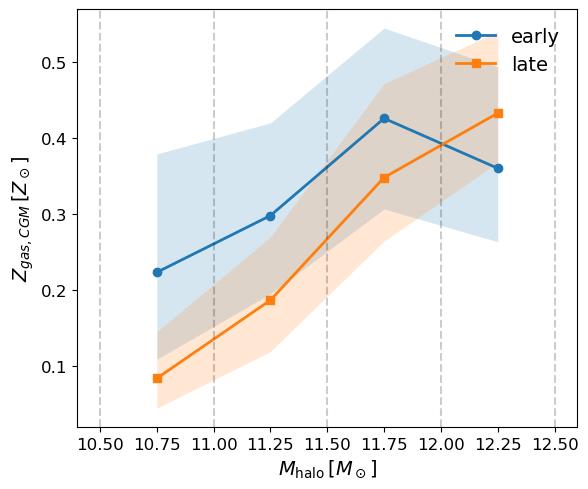}
}
\caption{The metallicity of the gas on galaxy and CGM scales. The left panel shows the metallicity in twice half stellar mass radius and the right panel presents the metallcity of the CGM gas. Symbols follow the same conventions as in Fig.~\ref{fig:halo_property}.}
\label{fig:metallicity}
\end{figure*}

Moreover, in the temperature range of $10^{4-7}\rm K$, metals dramatically enhance radiative cooling through metal-line cooling \citep[][]{sec33coolingSutherland1993,sec33coolingWiersma2009}. Higher metallicities thus reduce cooling times substantiallly and can alter both the thermal evolution of CGM gas and the subsequent star formation efficiency of the galaxy.

\begin{figure}
    \centering
    \includegraphics[width=\linewidth]{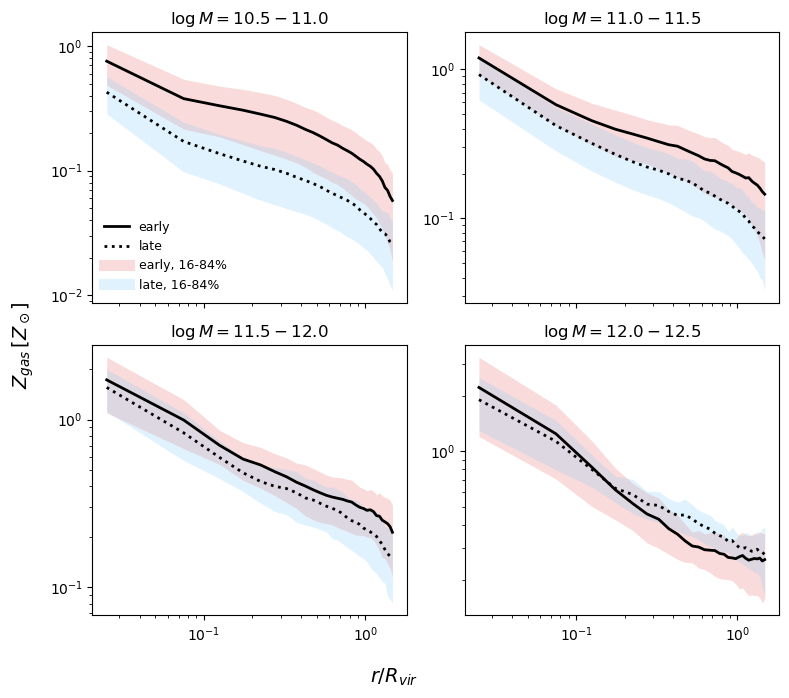}
    \caption{The radial distribution of the metallicity. The solid line shows the median of `early' halos while the dotted line shows the median of `late' halos. The shaded regions represent the 16-84th percentile.}
    \label{fig:zprofile}
\end{figure}

Given the important role of metals and in light of our previous results, it is reasonable to expect the assembly history of a halo would leave a measurable imprint on the metallicity of its CGM. Hence, we check the metallicity on both the galaxy (defined as the region within twice the stellar half mass radius) and CGM scales (Fig.~\ref{fig:metallicity}) , and assess whether galaxies with different assembly histories exhibit distinct metallicities profiles (Fig.~\ref{fig:zprofile}).

From the left panel of Fig.~\ref{fig:metallicity}, it is found that on galaxy scales, halos with earlier assembly histories exhibit systematically higher metallicities than their later counterparts. This is consistent with the results shown in Fig.~\ref{fig:halo_property}: `early' halos typically experience star formation over a longer period, undergoing more cycles of chemical enrichment, which in turn elevates their metlalicity on galaxy scale.

The situation slightly changes when we examine the metallicity on the CGM scales. The right panel of Fig.~\ref{fig:metallicity} shows that for halos with mass lower than $10^{12}\mathrm{M_\odot}$, the metallicity of CGM is higher in `early' halos. This reflects feedback-driven outflows that eject metal-rich gas from the ISM into the CGM. In these samples, outflows are primarily driven by supernovae rather than AGN. In contrast, for halos in the $10^{12-12.5} \mathrm{M_\odot}$ bin, the CGM of `early' halos exhibit a pronounced drop in the metallicity. To figure out the reason, we turn to Fig.~\ref{fig:zprofile}, which shows the radial distribution of the metallcity within the virial radius (in log space) . 

Across all halo mass bins, the radial metallicity gradients are generally negative and appear to show a slope change near $\sim 0.1R_{\rm vir}$, which is consistent with previous works \citep[][]{Suresh2015MNRAS.448..895S,Morgan2025ApJ...990...98M}.
Although the precise location of this change varies slightly across samples, it broadly corresponds to the transition between the ISM and the CGM. For halos with mass below $10^{12} \mathrm{M_\odot}$, the overall trend of the metallicity profile is similar between different assembly histories and becomes shallower over $\sim 0.1R_{vir}$-$0.5R_{vir}$ range. However, in the $10^{12-12.5} \mathrm{M_\odot}$ bin, despite the inner profiles are similar, `early' halos exhibit a steeper decline in metallicity over $\sim 0.1R_{vir}$-$0.5R_{vir}$ range relative to `late' halos . This indicates that galaxies in this mass bin enrich their central CGM less efficiently, contributing to the metallicity drop on CGM scale as we observed in Fig.~\ref{fig:metallicity}.

The reversal in the metallicity trend in `early' massive halos may be related to the results shown in Fig.~\ref{fig:halo_property}, together with the likely increasing importance of AGN kinetic feedback in this mass range. On the one hand, as shown in the right panel of Fig.~\ref{fig:halo_property}, galaxies in the massive halo regime are generally quenched at the present day, implying that `early' halos experience weaker ongoing stellar feedback than their later counterparts. This would reduce the supply of newly produced metals to the inner CGM \citep[][]{Morgan2025ApJ...990...98M}, and may therefore contribute to the steeper decline of the metallicity profile over the $\sim 0.1$-$0.5\,R_{\rm vir}$ range. On the other hand, AGN kinetic feedback may redistribute metal-rich gas from the inner regions into the outer CGM \citep[][]{Morgan2025ApJ...990...98M}, or even expel part of it beyond the virial radius into the surrounding environment \citep[][]{Suresh2015MNRAS.448..895S}. This could also help to explain the flatter metallicity profile at larger radii in the massive `early' halos.

Naturally, one might also expect the mergers could enrich the halo. However, as we will show in Section~\ref{subsec3.4:origin}, although `early' halos do not accrete more fresh, metal-poor gas than `late' halos, they acquire less gas through mergers or intergalactic transfer. With neither internal nor external supply replenishing the metal content, the CGM of the `early' halos in this mass bin inevitably experiences the metallicity drop we have observed.

\subsection{The origins of the CGM gas}
\label{subsec3.4:origin}

\begin{figure*}
    \includegraphics[width=0.95\textwidth]{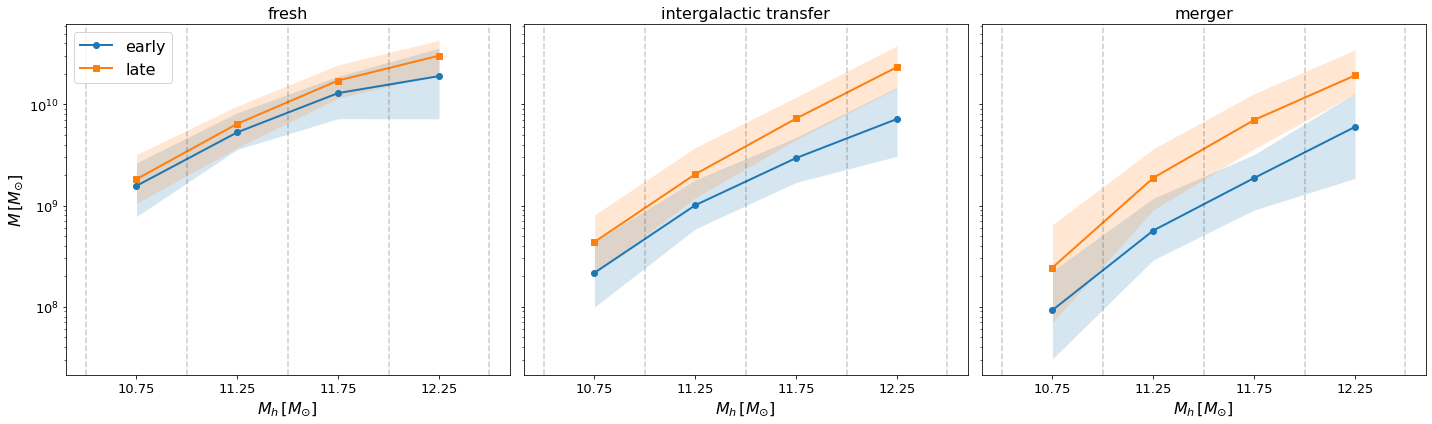}
    \caption{The mass comes from different origins. From left to right, the panels present the mass comes from fresh accretion, intergalactic tranfer, and merger-driven infall. Symbols follow the same conventions as in Fig.~\ref{fig:halo_property}.}
    \label{fig:morigin}
\end{figure*}

As discussed above, the gas residing in CGM has a diverse  origins, including feedback-driven outflows from the host galaxy, fresh accretion from the IGM, and merger-deriven infall carried by satellite or companion galaxies. These distinct sources generally possess markedly different physical properties.

For example, gas originating from host galaxy has undergone stellar nucleosynthesis and feedback heating, and therefore tends to be more metal-rich and to carry lower specific angular momentum under the influence of feedback \citep[][]{sec1haloshapeOppenheimer2008,sec33metaltracePeeples2014,sev34Zjupa2017}. In contrast, fresh accretion from the IGM is typically pristine or metal-poor, whereas gas brought in through mergers often enriches the primary halo because satellite galaxies have themselves undergone star formation, evolution, and mass loss, and such gas generally carries higher orbital angular momentum. Therefore, understanding how halos with different assembly histories modify their baryonic properties requires careful assessment of the relative contributions from varies origins.

In addition, halos with different assembly histories exhibit assembly bias: early-forming halos tend to reside in denser, more strongly correlated LSS environments. In such settings, they may be disadvantaged in the competition for cosmic gas supply, making it more difficult to replenish their gas reservoirs from the surrounding cosmic web or from other halos, for example `early' halos may be more susceptible to gas starvation near deep neighboring potential wells\citep[][]{Hahn2009MNRAS.398.1742H,Borzyszkowski2017MNRAS.469..594B,Voort2017MNRAS.466.3460V}.
Hence, it is interesting to see how the different origins of CGM gas are related with the assembly histories of halos.

In this work, we identify the origin of the CGM gas based on the criterion defined in \cite{AA2017}. In brief, we first identify all tracer particles attached to the CGM gas at $z=0$, and then track each tracer backward in time to determine the moment when it first enters the halo ($1^{st}$ accretion time). Prior to this $1^{st}$ accretion time, tracers that have spent less than 100 Myr in any other galaxy are classified as fresh accretion. If a tracer has spent more than 100 Myr within another galaxy, we further examine whether it resided in any galaxy for more than 100 Myr during the last 500 Myr before its first entry into the halo. Tracers satisfying this critera are identified as merger, whereas those that do not are identified as originating from intergalactic transfer.

Fig.~\ref{fig:morigin} compares the absolute masses contributed by the three channels. It is seen that the fresh accretion is similar in both `early' and `late' halos at lower halo masses, but begins to show some differences above $\sim 10^{12}\,\mathrm{M_\odot}$. By contrast, the contributions from the two external galaxy related channels (viz. intergalactic transfer and mergers) are systematically higher in the `late' halos.
This could also contributes the metallicity drop we found in Section~\ref{subsec3.3:metal} for `early' halos in the $10^{12-12.5} \mathrm{M_\odot}$ bin: with little internal enrichment at late times and no significant external supplement of metal-rich gas, consequently their CGM becomes dominated by fresh, metal-poor gas, which effectively dilutes the CGM metallicity.

Taken together, these results suggest that differences in assembly history largely reflect differences in environment. Early-forming halos, often located near more massive neighbors, are inefficient at capturing gas from other galaxies — whether through intergalactic transfer or merger - driven infall. However, despite their environmental disadvantage, the impact on the amount of fresh IGM accretion is relatively modest.

\subsection{CGM kinetics: angular momentum and mass flows}
\label{subsec3.5:kinetics}

\begin{figure*}
    \includegraphics[width=0.9\textwidth]{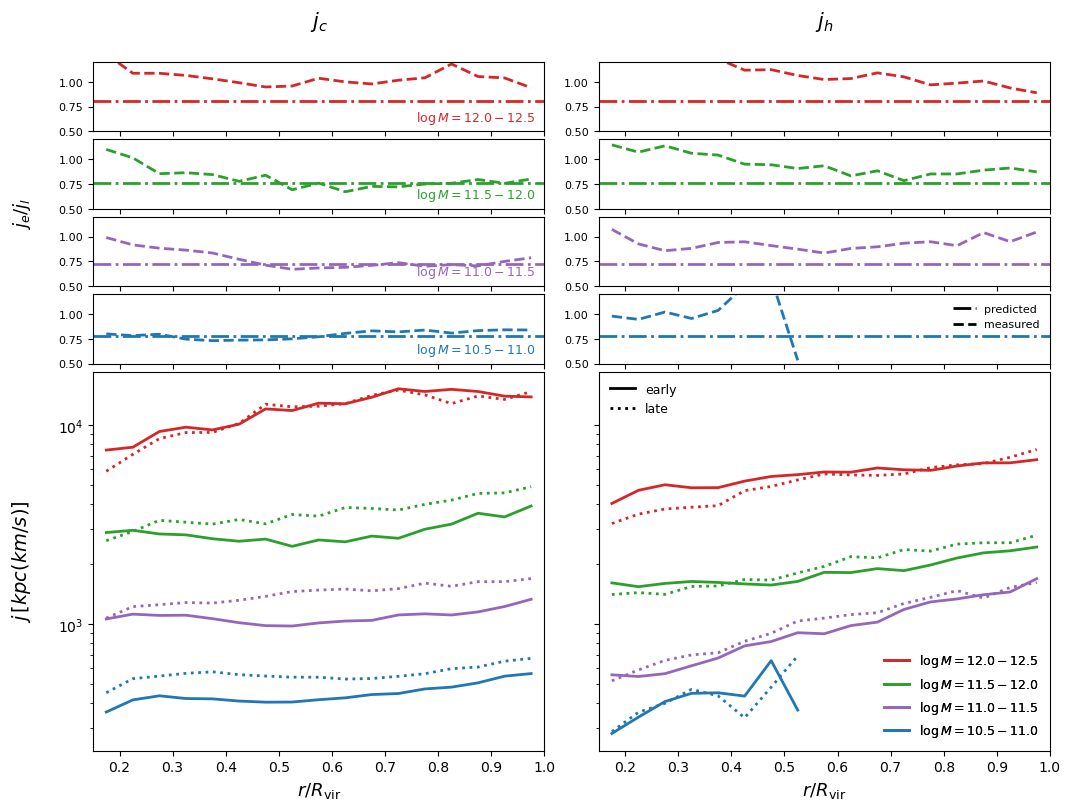}
    \includegraphics[width=0.9\textwidth]{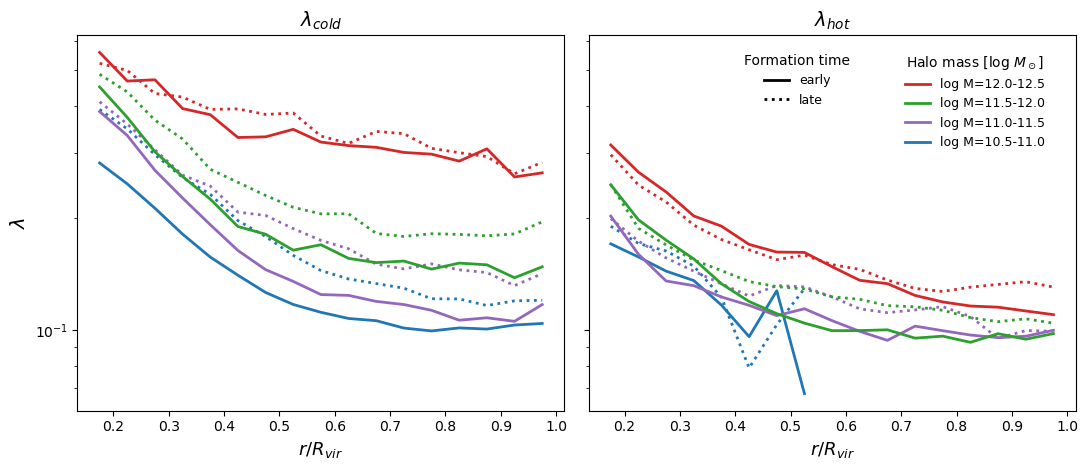}
\caption{The median radial profiles of the specific angular momentum and Bullock spin parameter for halos with different assembly histories. The upper row shows the specific angular momentum, together with the comparison between the predicted and measured early-to-late ratios, while the second row shows the Bullock spin parameter. In each row, the left panel corresponds to the cold phase and the right panel to the hot phase. The colors and line styles follow the same convention as in Fig.~\ref{fig:rhoprofile}.For the eight subpanels in the upper row, the dash-dotted horizontal lines indicate the early-to-late specific-angular-momentum ratio predicted from the scaling relation, while the dashed lines indicate the median early-to-late ratio measured directly from the simulation. }
\label{fig:jandlambda}
\end{figure*}

In this part we investigate the kinematic properties of the CGM. Specifically, we examine the specific angular momentum and spin parameters of the cold and hot gas phases, and the mass fluxes.

Fig.~\ref{fig:jandlambda} presents the median radial profiles of the specific angular momentum and spin parameter for halos with different assembly histories. Within each radial shell, the specific angular momentum are computed following equation~\ref{eq:j} and spin parameter defined following \cite{spinparameterBullock2001}, as given in equation~\ref{eq:lambda}.

\begin{equation}
    j=\frac{J}{M}=\frac{\sum (r_i \times v_i) m_i}{\sum m_i}
    \label{eq:j}
\end{equation}

\begin{equation}
    \lambda=\frac{j_{shell}}{\sqrt{2}RV}
    \label{eq:lambda}
\end{equation}

where i represents the gas cell within the shell and $V=\sqrt{\frac{GM}{R}}$, the M is the total mass enclosed within that radius.

To quantify the separation between the `early' and `late' median trends, we performed an additional permutation-based statistical test, a brief description of which is provided in Appendix~\ref{appendix:perm}. We adopted it here rather than overplotting shaded regions for every curve, because in practice the large number of lines in these figures would make the shaded regions visually crowded and difficult to interpret. The results are listed in Table~\ref{tab:jandlambda}. Simply, smaller p-values indicate a more significant difference between the `early' and `late' trends.

\begin{table}[]
    \centering
    \begin{tabular}{c|c|c|c|c}
        \hline
         $M_{h}$ & $j_c$ & $j_h$ & $\lambda_{cold}$ & $\lambda_{hot}$ \\
         \hline
        $10^{10.5-11}$ & 0.0002 & 0.4273 & 0.0002 & 0.3343 \\
        $10^{11-11.5}$ & 0.0002 & 0.0226 & 0.0002 & 0.0072 \\
        $10^{11.5-12}$ & 0.0038 & 0.1186 & 0.0032 & 0.0520 \\
        $10^{12-12.5}$ & 0.9636 & 0.4427 & 0.8194 & 0.6017 \\
        \hline
    \end{tabular}
    \caption{Permutation-test $p$-values for the separations between the `early' and `late' median radial profiles of the specific angular momentum and spin parameter. Each row corresponds to a halo-mass bin, and each column to one of the profile quantities shown in Fig.~\ref{fig:jandlambda}.}
    \label{tab:jandlambda}
\end{table}

It is found that, compared with `early' halos, the cold phase CGM in `late' halos typically carries higher specific angular momentum and has a higher degree of rotational support. The hot gas, in contrast, shows little difference across assembly histories. A plausible explanation is that cold CGM gas more likely retains the orbital angular momentum it possessed upon infall and that its specific angular momentum at $z=0$ is established primarily around the formation time, whereas the more diffuse, high temperature gas loses memory of its initial kinematic state.

To illustrate this possibility, we use the median formation redshift to estimate the expected ratio of specific angular momentum between `early' and `late' halos based on the well-known proportional relation: $j \propto M_h^{2/3} / \sqrt{1+z}$ \citep[][]{sec1araa2023}. At the same time, we also computed the corresponding ratio measured directly from the median values in the simulation; the result is shown in the top sub-panel of Fig.~\ref{fig:jandlambda}.

It is found that for halos below $10^{12}\mathrm{M_\odot}$, the specific angular momentum ratio of the cold CGM gas agrees well with the theoretical prediction, in support of our hypothesis. However, in the $10^{12-12.5}\mathrm{M_\odot}$ bin, the difference between `early' and `late' halos becomes small, and the actual value do not match the predicted ones well. This can be understood in light of the results of \cite{Dekel2006}, which indicate that halos in this mass range are primarily fed by hot mode accretion. So most of the cold CGM gas in these systems is likely produced by cooling of recycled material rather than by direct cosmological inflow, making the simple theoretical estimate less applicable.

We note that the scaling relation used here is, strictly speaking, defined for the global specific angular momentum rather than for the radial profile at each radius. Nevertheless, Fig.~\ref{fig:jandlambda} shows that the gas specific-angular-momentum profiles vary only modestly with radius over the range considered. For this reason, we use the relation only as a heuristic reference for the approximate early-to-late ratio across radius, rather than as a rigorous local prediction. However, the origin of the CGM angular momentum remains an open question, and the interpretation offered here is only one plausible scenario. A more complete understanding will require further investigation.

Finally, we examine the mass flux across different radii within the CGM. Following \cite{TNG50Nelson}, the radial gas mass flux is defined in equation~\ref{eq:mdot}, with the outward direction from the galaxy center taken to be positive; thus, positive values indicate outflows, while negative values correspond to inflows. 

\begin{equation}
    \dot{M}=\frac{1}{\Delta r}\sum_{\substack{i=0 \\ |r_i - r_0| \leq \Delta r / 2}}^n (\frac{\vec{v}_i\cdot\vec{r}_i}{|r_i|}m_i)
    \label{eq:mdot}
\end{equation}

\begin{figure*}
    \centering
    \includegraphics[width=0.95\linewidth]{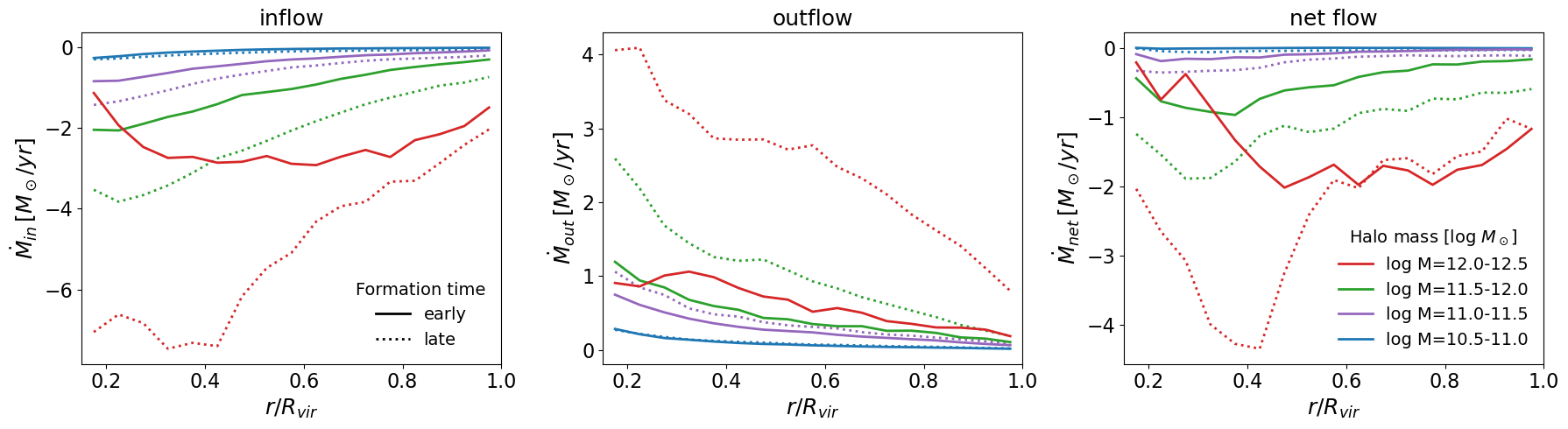}
    \caption{The mass flux at different radii. From left to right, panels show the inflow, outflow, and net mass flux, respectively. Positive values indicate outflows, while negative values indicate inflows. Colors and line styles follow the same convention as in Fig.~\ref{fig:rhoprofile}.}
    \label{fig:massflow}
\end{figure*}

The results in Fig.~\ref{fig:massflow} and Table~\ref{tab:massflow} show that both inflow and outflow rates tend to be lower in `early' halo across all radii compared to their `late' counterparts, and the net mass flow is inflow for all halos. This suppressed activity implies that the internal environments of `early' galaxies are generally more quiescent, likely reflecting a state of virial equilibrium or the transition into a preventative feedback regime. 
In contrast, `late' halos exhibit more vigorous gas exchange, indicative of ongoing rapid assembly. This behavior is expected. According to \cite{Wechsler2002ApJ...568...52W}, the mass growth of dark matter halo can be described by a simple parametric form: $M_h(z)=M_{h0}e^{-\alpha z}$, where $z$ is redshift. From this formula, the $\dot{M_h}$ is proportional to the growth parameter $\alpha$.
For halos with comparable masses at $z=0$, `late' halos are characterized by larger values of $\alpha$. Therefore, `late' halos would exhibit higher mass accretion rates at $z=0$ than `early' halos (at this time the accretion rate is proportion to $\alpha$). Furthermore, as will be shown in Section~\ref{sec4:discussion}, halo mass growth in `late' halos is more efficient at bringing in gas, which in turn can drive stronger gas inflows into the CGM.
As a consequence, this abundant fuel supply drives higher star formation rates in the central galaxies of `late' halos, which we have seen in Fig.~\ref{fig:halo_property}. Since galactic outflows are primarily feedback-driven, the elevated SFR in `late' halos naturally results in stronger, more mass-loaded outflows\citep[][]{TNG50Nelson}.

\begin{table}[]
    \centering
    \begin{tabular}{c|c|c|c}
        \hline
         $M_{h}$ & $\dot{M}_{in}$ & $\dot{M}_{out}$ & $\dot{M}_{net}$ \\
         \hline
        $10^{10.5-11}$ & 0.0002 & 0.0172 & 0.0002 \\
        $10^{11-11.5}$ & 0.0002 & 0.0002 & 0.0002 \\
        $10^{11.5-12}$ & 0.0002 & 0.0002 & 0.0002 \\
        $10^{12-12.5}$ & 0.0002 & 0.0002 & 0.0352 \\
        \hline
    \end{tabular}
    \caption{Permutation-test $p$-values for the separations between the `early' and `late' median radial profiles of the mass flux. Each row corresponds to a halo-mass bin, and each column to one of the profile quantities shown in Fig.~\ref{fig:massflow}. }
    \label{tab:massflow}
\end{table}

\section{Discussion}
\label{sec4:discussion}

In Section~\ref{sec3:result}, we have shown that within a narrow halo mass range, halos with different formation times can exhibit differences in their CGM properties, such as gas mass, metallicity, specific angular momentum and gas flow rates. This indicates that assembly history is a non-negligible factor in shaping the CGM, a conclusion that has also been emphasized in previous observational and simulation studies\citep[][]{Lim2016MNRAS.455..499L,Tojeiro2017MNRAS.470.3720T,Zehavi2018ApJ...853...84Z,Bordoloi2018ApJ...864..132B,D19,Montero2021MNRAS.508..940M,D21}. Discussions of the impact of assembly history on the CGM generally focus on two aspects: differences in the cumulative effects of internal feedback induced by different formation times, and differences in the large-scale external environments traced by formation time. Previous works such as \cite{D19} and \cite{D21} have provided detailed investigations of the former using the EAGLE simulations. However, it is increasingly clear that the external environment also plays a crucial role in regulating CGM properties\citep[][]{Yoon2013ApJ...772L..29Y,Burchett2016ApJ...832..124B,Wang2022MNRAS.509.3148W}.

\begin{figure}
    \includegraphics[width=0.45\textwidth]{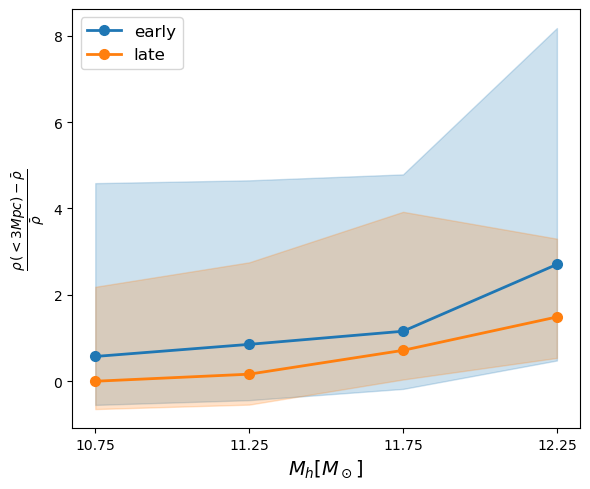}
    \includegraphics[width=0.45\textwidth]{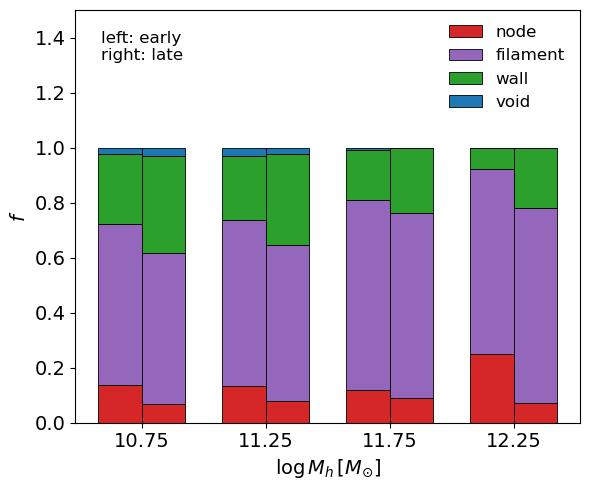}
    \caption{Different environments of `early' and `late' halos in each mass bin. Top-panel: Normalized over-density. Symbols show the medians, the shaded regions indicate the 16-84th percentile ranges of the full distributions. Bottom-panel: Fraction of halos residing in different cosmic-web types (node, filament, wall, and void). Within each halo-mass bin, the left and right stacked bars correspond to the `early' and the `late' populations, respectively.}
    \label{fig:overdensity}
\end{figure}

Fig.~\ref{fig:overdensity} demonstrates the different environments of the `early' and `late' halo populations across the four halo-mass bins. The top panel shows the normalized over-density around each galaxy, defined as $\delta \equiv (\rho(<3\,{\rm Mpc}) - \bar{\rho})/\bar{\rho}$, where $\rho(<3\,{\rm Mpc})$ is the mass density computed within a sphere of radius 3\,Mpc centered on each galaxy, and $\bar{\rho}$ is the mean mass density of all subhalos in the simulation box. The bottom panel shows the fractions of four LSS types (i.e.node, filament, wall, void) for our `early' and `late' halos samples. `Early' halos tend to live in denser environments, as predicted by the assembly bias theory\citep[][]{Gao2005}. As shown, their median local densities are systematically higher than those of the `late' population, and they also exhibit a higher fraction of node environments.
Taken together, these results evident our `early' and `late' halo samples tend to reside in different LSS environments. This, in turn, supports the idea that environmental differences are likely to contribute to the CGM trends discussed above, although they are unlikely to be the sole driver, and internal processes such as the cumulative impact of feedback may also play an important role.\citep[][]{D19,D21}

Specifically, early-forming halos tend to reside in more clustered environments and are more likely to be surrounded by massive neighboring halos. Gas in such environments are generally hotter and less easily retained. Satellite halos in these regions are more susceptible to environmental processes such as ram-pressure stripping and tidal interactions, which can efficiently remove their gas reservoirs prior to accretion \citep[][, so called `pre-processings']{Fujita2004PASJ...56...29F,Vijayaraghavan2013MNRAS.435.2713V} . As a result of these pre-processings, satellites falling into `early' halos are expected to be gas-poor and characterized as dry mergers, which contribute a smaller proportion of gas to the CGM.

\begin{figure}
    \centering
    \includegraphics[width=0.95\linewidth]{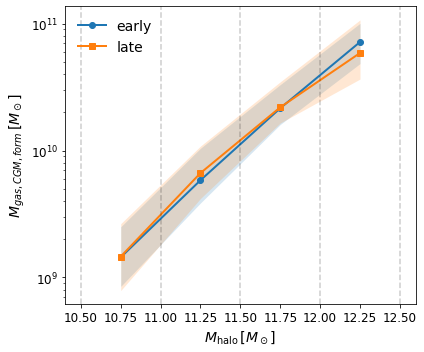}
    \caption{The gas mass in the CGM of each halo sample at its formation time. Symbols follow the same conventions as in Fig.~\ref{fig:halo_property}.}
    \label{fig:mcgmgasform}
\end{figure}

This picture explains the trends illustrated in Fig.~\ref{fig:morigin}. The enhanced contribution of mergers and intergalactic transfer to the CGM in `late' halos motivates us to explore at which stages these differences are established? To address this, in Fig.~\ref{fig:mcgmgasform}, we first compare the CGM gas masses of `early' and `late' halos at their formation times. 
It is found that at the formation time, the CGM gas masses of the `early' and `late' are nearly identical, this suggests that the disparities in the CGM gas mass and other CGM properties observed at $z=0$ are established after the formation time. 

\begin{figure}
    \centering
    \includegraphics[width=0.95\linewidth]{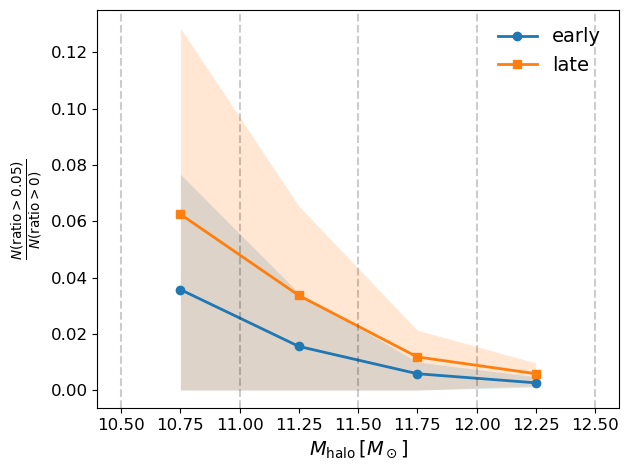}
    \caption{Fraction of mergers in which the gas mass of the infalling satellite exceeds 5\% of the gas mass of the central galaxy on the main branch after the formation time. Symbols follow the same conventions as in Fig.~\ref{fig:halo_property}.}
    \label{fig:wetordry}
\end{figure}

This raises the question of whether these differences originate from the different merger rates after the formation time. However, we found that `early' halos have experienced similar or even slightly more merger events than the `late' halos by approximately 0.15-0.3 dex. This indicates that the differences of CGM gas can not explained by the merger rate merely. These differences could be originated from the different gas mass ratios of mergers, i.e., dry merger or wet merger.
Motivated by the picture of `pre-processing', we define a criterion to quantify the quality of mergers: a merger is considered gas-rich if the gas mass of the accreted satellite exceeds 5\% of the gas mass of the central halo along the main progenitor branch. 
We then compute, for mergers occurring after the formation time, the fraction of mergers satisfying this criterion relative to the total number of mergers, and present the results in Fig.~\ref{fig:wetordry}. This fraction is generally higher in `late' halos than in `early' halos across the full halo-mass range considered, with the difference being most pronounced at lower masses and becoming much smaller toward the highest-mass bin. This demonstrates that `late' halos experience more `wet' mergers which contribute effectively to the growth of the CGM gas mass. In contrast, mergers in `early' halos predominantly increase halo mass without providing a comparable gas supply, ultimately leading to the observed differences in CGM gas mass at $z=0$.

\begin{figure}
    \centering
    \includegraphics[width=0.95\linewidth]{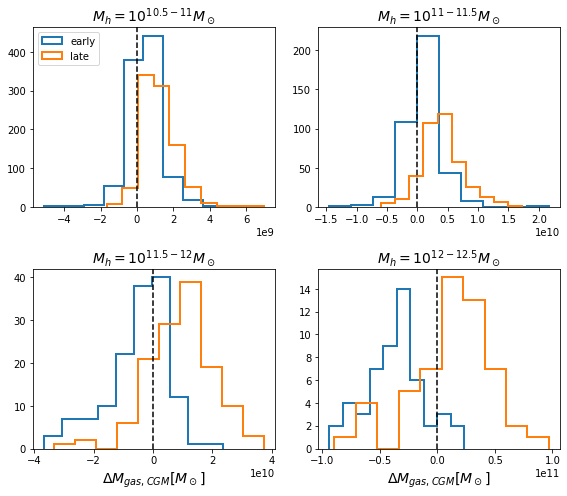}
    \caption{The difference in CGM gas mass between $z=0$ and the formation time. Blue and orange histograms denote `early' and `late' halos, respectively. The vertical dashed line separates net mass growth from mass loss.}
    \label{fig:deltaMgascgm}
\end{figure}

Furthermore, we compare the difference between the CGM gas mass at $z=0$ and that at the formation time. As shown in Fig.~\ref{fig:deltaMgascgm}, it is found that for halos in the mass range $10^{11.5-12.5} M_\odot$, `early' halos exhibit a pronounced decrease in CGM gas mass. This behavior could be likely associated with the enhanced impact of black hole feedback in this mass regime (beyond the scope of our paper), which has been discussed by many works\citep[e.g.,][]{D19,Oppenheimer2020MNRAS.491.2939O,sec1smbhZinger2020,Frosst2025MNRAS.537.3543F}.

Based on our current results, we can outline a brief picture to interpret the differences identified in Section~\ref{sec3:result}. For early-forming halos, two competing effects are at play. On the one hand, because they tend to reside in denser environments, in which surrounding low-mass halos are more likely to experience substantial pre-processing before being accreted by `early' halos. As a result, accreted satellite galaxies contain less gas (dry mergers) as shown in Fig.~\ref{fig:morigin} and Fig.~\ref{fig:wetordry}. On the other hand, early-forming halos have experienced longer periods of star formation, leading to the buildup of higher stellar masses and metal abundances. This enhanced metal production enriches the CGM, resulting in elevated metallicities on both galactic and circumgalactic scales, particularly in the inner region (Fig.~\ref{fig:halo_property} and Fig.~\ref{fig:metallicity}). However, for halos in the mass range $10^{12-12.5}M_\odot$, AGN feedback would become sufficiently strong to efficiently expel CGM gas and quench star formation (Fig.~\ref{fig:deltaMgascgm} and Fig.~\ref{fig:halo_property}). The combined effects of enhanced gas expulsion and reduced gas replenishment substantially amplifying the CGM gas mass differences between `early' and `late' halos (Fig.~\ref{fig:cgm_thermo_property}) and ultimately leading to a pronounced drop in CGM-scale metallicity at the massive end for `early' halos (Fig.~\ref{fig:metallicity}).

A complementary picture can be drawn for the late-forming halos. On average, these halos tend to reside in somewhat less dense environments, so the galaxies accreted onto them are likely to have experienced less environmental processing prior to infall. As a result, satellite accretion remains relatively gas-rich, yielding wet mergers and larger contributions from the externally supplied channels. At the same time, because late-forming halos assemble more recently, their central galaxies generally have lower stellar masses and somewhat lower metal production at fixed halo mass. This tends to reduce the degree of metal enrichment, especially in the inner halo. However, the larger CGM gas reservoir maintained by continued gas supply, together with less efficient gas removal in the massive halos, helps sustain higher CGM gas masses and prevents the outer-halo metallicity from dropping as strongly as in the early-forming systems. 

Although we interpret the differences between the `early' and `late' halos from the environmental prospect, we emphasize that this is unlikely to be the only factor at work. In particular, for halos in the $10^{11.5-12.5}\,\mathrm{M_\odot}$ range, internal feedback processes  also play a non-negligible role. Disentangling the contributions from the environment and internal feedback to the observed halo-to-halo differences require further investigation in the future.

\section{Summary}
\label{sec5:summary}

In this work, we employed TNG50-1, which is the highest resolution simulation in the IllustrisTNG suite, to investigate how different assembly histories affect both galaxy properties and the CGM across four halo-mass bins spanning $10^{10.5-12.5}\mathrm{M_\odot}$. Our main findings are summarized as follows:
\begin{itemize}
    \item Early-forming halos tend to host galaxies with higher stellar masses, but with lower baryon fraction and present-day specific star formation rate. Specifically, `early' halos with $M_h=10^{12-12.5}\rm M_\odot$ are  predominantly quenched, with about 70\% of the samples exhibiting sSFR $<10^{-11}\rm yr^{-1}$ (Fig.~\ref{fig:halo_property}).
    \item The CGM gas mass in early-forming halos is significantly lower than in late-forming halos (Fig.~\ref{fig:cgm_thermo_property}). Furthermore, assembly history primarily affects the amount of CGM gas rather than its spatial distribution (Fig.~\ref{fig:rhoprofile}). 
    \item Early-forming halos exhibit higher metallicities on galaxy scales. In the CGM, this trend persists below $10^{12}\mathrm{M_\odot}$ but reverses in the $10^{12-12.5}\mathrm{M_\odot}$ bin: early-forming halos show a marked metallicity drop (Fig.~\ref{fig:metallicity}). This is because `early' halos in this mass bin are typically quenched at present day, suppressing the metal enrichment of CGM in the inner region. And at the same time, their CGM would be further diluted by fresh accretion while receiving less replenishment of metal-enriched gas from mergers or intergalactic transfer (Fig.~\ref{fig:morigin}).
    \item Early-forming halos tend to reside in more clustered environments (Fig.~\ref{fig:overdensity}), but this has only a modest impact on the amount of fresh accretion. The major differences arise instead from gas supplied by merging satellite: late-forming halos systematically acquire more gas from both mergers and intergalactic transfer (Fig.~\ref{fig:morigin}). 
    \item At the formation time, i.e. the time at which half the mass is assembled, the CGM gas masses of `early' and `late' halos are similar, the differences observed at present-day arise primarily from evolutionary processes during the second half of assembly (Fig.~\ref{fig:mcgmgasform}). It is found that `late' halos experience a higher fraction of wet mergers after formation time. In contrast, mergers occurring in `early' halos after the formation time tend to contribute predominantly to halo mass growth while providing relatively little gas to the CGM (Fig.~\ref{fig:wetordry}). This divergence is likely linked to the distinct environments in which `early' and `late' halos reside.
    \item For halos below $10^{12}\mathrm{M_\odot}$, the cold gas in late-forming halos carries higher specific angular momentum and has a higher degree of rotational support, while the hot gas shows little dependence on assembly history. In the $10^{12-12.5}\mathrm{M_\odot}$ bin, both cold and hot phases show no significant differences between early and late assembly histories (Fig.~\ref{fig:jandlambda}).
    \item Across all halo mass bins, early-forming halos exhibit systematically weaker gas mass inflows and outflows at all radii, reflecting a more quiescent state of present-day galaxy activity (Fig.~\ref{fig:massflow}).
\end{itemize}

This work mainly focuses on investigating whether present-day CGM properties differ among halos with different formation time. Our results clearly demonstrate that, within the TNG framework, assembly history exerts a non-negligible impact on both galactic and circumgalactic properties across $10^{10.5-12.5}\mathrm{M_\odot}$ range. And we have verified that the corresponding trends for the intermediate halos lie between those of the `early' and `late' populations, indicating that the differences reported here reflect a systematic dependence on formation time rather than these two sub-samples being outliers of the full distribution. These findings would provide a useful and conceivable perspective for interpreting the diversity of CGM and galaxy properties.

Our results mainly indicate that environmental effects provide one plausible explanation for the difference of CGM in halos with different formation times. 
Owing to limitations in the available data, we are unable to follow the baryon cycling processes that occur after gas accretion into galaxies, which are likely important for a more complete understanding of variations in metal enrichment and phase transitions among different gas components. A more comprehensive understanding of these processes is essential for fully characterizing the origin of CGM diversity and we would explore this in more detail in the future by employing zoom-in simulations.

\begin{acknowledgments}
This work is supported by NSFC (No. 12595314, 12533007, 12547104, 12233005), the National Key Research and Development Program (No. 2022YFA1602903, No. 2023YFB3002502), the science research grants from the China Manned Space project with NO. CMS-CSST-2025-A10. We sincerely thank D. Nelson for providing additional tracer data, which greatly facilitated the completion of this work. YYZ is grateful to Jiafeng Lu, Tingting Tian and Yiheng Wang for insightful discussions. The analysis presented in this article were carried out on the SilkRiver Supercomputer of Zhejiang University, located at the Zhejiang University Information Center. Finally, we thank the referee for the constructive comments and suggestions, which have improved this paper.
\end{acknowledgments}

\bibliography{reference}{}

@ARTICLE{sec1araa2023,
       author = {{Faucher-Gigu{\`e}re}, Claude-Andr{\'e} and {Oh}, S. Peng},
        title = "{Key Physical Processes in the Circumgalactic Medium}",
      journal = {\araa},
         year = 2023,
        month = aug,
       volume = {61},
        pages = {131-195},
          doi = {10.1146/annurev-astro-052920-125203},
archivePrefix = {arXiv},
       eprint = {2301.10253},
 primaryClass = {astro-ph.GA},
       adsurl = {https://ui.adsabs.harvard.edu/abs/2023ARA&A..61..131F}
}

@ARTICLE{sec1araa2017,
       author = {{Tumlinson}, Jason and {Peeples}, Molly S. and {Werk}, Jessica K.},
        title = "{The Circumgalactic Medium}",
      journal = {\araa},
         year = 2017,
        month = aug,
       volume = {55},
       number = {1},
        pages = {389-432},
          doi = {10.1146/annurev-astro-091916-055240},
archivePrefix = {arXiv},
       eprint = {1709.09180},
 primaryClass = {astro-ph.GA},
       adsurl = {https://ui.adsabs.harvard.edu/abs/2017ARA&A..55..389T}
}

@ARTICLE{sec1galex,
       author = {{Martin}, D. Christopher and {Fanson}, James and {Schiminovich}, David and {Morrissey}, Patrick and {Friedman}, Peter G. and {Barlow}, Tom A. and {Conrow}, Tim and {Grange}, Robert and {Jelinsky}, Patrick N. and {Milliard}, Bruno and {Siegmund}, Oswald H.~W. and {Bianchi}, Luciana and {Byun}, Yong-Ik and {Donas}, Jose and {Forster}, Karl and {Heckman}, Timothy M. and {Lee}, Young-Wook and {Madore}, Barry F. and {Malina}, Roger F. and {Neff}, Susan G. and {Rich}, R. Michael and {Small}, Todd and {Surber}, Frank and {Szalay}, Alex S. and {Welsh}, Barry and {Wyder}, Ted K.},
        title = "{The Galaxy Evolution Explorer: A Space Ultraviolet Survey Mission}",
      journal = {\apjl},
         year = 2005,
        month = jan,
       volume = {619},
       number = {1},
        pages = {L1-L6},
          doi = {10.1086/426387},
archivePrefix = {arXiv},
       eprint = {astro-ph/0411302},
 primaryClass = {astro-ph},
       adsurl = {https://ui.adsabs.harvard.edu/abs/2005ApJ...619L...1M}
}

@ARTICLE{sec1sdss,
       author = {{York}, Donald G. and {Adelman}, J. and {Anderson}, Jr., John E. and {Anderson}, Scott F. and {Annis}, James and {Bahcall}, Neta A. and {Bakken}, J.~A. and {Barkhouser}, Robert and {Bastian}, Steven and {Berman}, Eileen and {Boroski}, William N. and {Bracker}, Steve and {Briegel}, Charlie and {Briggs}, John W. and {Brinkmann}, J. and {Brunner}, Robert and {Burles}, Scott and {Carey}, Larry and {Carr}, Michael A. and {Castander}, Francisco J. and {Chen}, Bing and {Colestock}, Patrick L. and {Connolly}, A.~J. and {Crocker}, J.~H. and {Csabai}, Istv{\'a}n and {Czarapata}, Paul C. and {Davis}, John Eric and {Doi}, Mamoru and {Dombeck}, Tom and {Eisenstein}, Daniel and {Ellman}, Nancy and {Elms}, Brian R. and {Evans}, Michael L. and {Fan}, Xiaohui and {Federwitz}, Glenn R. and {Fiscelli}, Larry and {Friedman}, Scott and {Frieman}, Joshua A. and {Fukugita}, Masataka and {Gillespie}, Bruce and {Gunn}, James E. and {Gurbani}, Vijay K. and {de Haas}, Ernst and {Haldeman}, Merle and {Harris}, Frederick H. and {Hayes}, J. and {Heckman}, Timothy M. and {Hennessy}, G.~S. and {Hindsley}, Robert B. and {Holm}, Scott and {Holmgren}, Donald J. and {Huang}, Chi-hao and {Hull}, Charles and {Husby}, Don and {Ichikawa}, Shin-Ichi and {Ichikawa}, Takashi and {Ivezi{\'c}}, {\v{Z}}eljko and {Kent}, Stephen and {Kim}, Rita S.~J. and {Kinney}, E. and {Klaene}, Mark and {Kleinman}, A.~N. and {Kleinman}, S. and {Knapp}, G.~R. and {Korienek}, John and {Kron}, Richard G. and {Kunszt}, Peter Z. and {Lamb}, D.~Q. and {Lee}, B. and {Leger}, R. French and {Limmongkol}, Siriluk and {Lindenmeyer}, Carl and {Long}, Daniel C. and {Loomis}, Craig and {Loveday}, Jon and {Lucinio}, Rich and {Lupton}, Robert H. and {MacKinnon}, Bryan and {Mannery}, Edward J. and {Mantsch}, P.~M. and {Margon}, Bruce and {McGehee}, Peregrine and {McKay}, Timothy A. and {Meiksin}, Avery and {Merelli}, Aronne and {Monet}, David G. and {Munn}, Jeffrey A. and {Narayanan}, Vijay K. and {Nash}, Thomas and {Neilsen}, Eric and {Neswold}, Rich and {Newberg}, Heidi Jo and {Nichol}, R.~C. and {Nicinski}, Tom and {Nonino}, Mario and {Okada}, Norio and {Okamura}, Sadanori and {Ostriker}, Jeremiah P. and {Owen}, Russell and {Pauls}, A. George and {Peoples}, John and {Peterson}, R.~L. and {Petravick}, Donald and {Pier}, Jeffrey R. and {Pope}, Adrian and {Pordes}, Ruth and {Prosapio}, Angela and {Rechenmacher}, Ron and {Quinn}, Thomas R. and {Richards}, Gordon T. and {Richmond}, Michael W. and {Rivetta}, Claudio H. and {Rockosi}, Constance M. and {Ruthmansdorfer}, Kurt and {Sandford}, Dale and {Schlegel}, David J. and {Schneider}, Donald P. and {Sekiguchi}, Maki and {Sergey}, Gary and {Shimasaku}, Kazuhiro and {Siegmund}, Walter A. and {Smee}, Stephen and {Smith}, J. Allyn and {Snedden}, S. and {Stone}, R. and {Stoughton}, Chris and {Strauss}, Michael A. and {Stubbs}, Christopher and {SubbaRao}, Mark and {Szalay}, Alexander S. and {Szapudi}, Istvan and {Szokoly}, Gyula P. and {Thakar}, Anirudda R. and {Tremonti}, Christy and {Tucker}, Douglas L. and {Uomoto}, Alan and {Vanden Berk}, Dan and {Vogeley}, Michael S. and {Waddell}, Patrick and {Wang}, Shu-i. and {Watanabe}, Masaru and {Weinberg}, David H. and {Yanny}, Brian and {Yasuda}, Naoki and {SDSS Collaboration}},
        title = "{The Sloan Digital Sky Survey: Technical Summary}",
      journal = {\aj},
         year = 2000,
        month = sep,
       volume = {120},
       number = {3},
        pages = {1579-1587},
          doi = {10.1086/301513},
archivePrefix = {arXiv},
       eprint = {astro-ph/0006396},
 primaryClass = {astro-ph},
       adsurl = {https://ui.adsabs.harvard.edu/abs/2000AJ....120.1579Y}
}

@ARTICLE{sec1cos,
       author = {{Green}, James C. and {Froning}, Cynthia S. and {Osterman}, Steve and {Ebbets}, Dennis and {Heap}, Sara H. and {Leitherer}, Claus and {Linsky}, Jeffrey L. and {Savage}, Blair D. and {Sembach}, Kenneth and {Shull}, J. Michael and {Siegmund}, Oswald H.~W. and {Snow}, Theodore P. and {Spencer}, John and {Stern}, S. Alan and {Stocke}, John and {Welsh}, Barry and {B{\'e}land}, St{\'e}phane and {Burgh}, Eric B. and {Danforth}, Charles and {France}, Kevin and {Keeney}, Brian and {McPhate}, Jason and {Penton}, Steven V. and {Andrews}, John and {Brownsberger}, Kenneth and {Morse}, Jon and {Wilkinson}, Erik},
        title = "{The Cosmic Origins Spectrograph}",
      journal = {\apj},
         year = 2012,
        month = jan,
       volume = {744},
       number = {1},
          eid = {60},
        pages = {60},
          doi = {10.1088/0004-637X/744/1/6010.1086/141956},
archivePrefix = {arXiv},
       eprint = {1110.0462},
 primaryClass = {astro-ph.IM},
       adsurl = {https://ui.adsabs.harvard.edu/abs/2012ApJ...744...60G}
}

@ARTICLE{sec1cxo,
       author = {{Weisskopf}, M.~C. and {Brinkman}, B. and {Canizares}, C. and {Garmire}, G. and {Murray}, S. and {Van Speybroeck}, L.~P.},
        title = "{An Overview of the Performance and Scientific Results from the Chandra X-Ray Observatory}",
      journal = {\pasp},
         year = 2002,
        month = jan,
       volume = {114},
       number = {791},
        pages = {1-24},
          doi = {10.1086/338108},
archivePrefix = {arXiv},
       eprint = {astro-ph/0110308},
 primaryClass = {astro-ph},
       adsurl = {https://ui.adsabs.harvard.edu/abs/2002PASP..114....1W}
}

@ARTICLE{sec1xmmnewton,
       author = {{Jansen}, F. and {Lumb}, D. and {Altieri}, B. and {Clavel}, J. and {Ehle}, M. and {Erd}, C. and {Gabriel}, C. and {Guainazzi}, M. and {Gondoin}, P. and {Much}, R. and {Munoz}, R. and {Santos}, M. and {Schartel}, N. and {Texier}, D. and {Vacanti}, G.},
        title = "{XMM-Newton observatory. I. The spacecraft and operations}",
      journal = {\aap},
         year = 2001,
        month = jan,
       volume = {365},
        pages = {L1-L6},
          doi = {10.1051/0004-6361:20000036},
       adsurl = {https://ui.adsabs.harvard.edu/abs/2001A&A...365L...1J}
}

@ARTICLE{sec1meaningfulsampleCookesy2010,
       author = {{Cooksey}, Kathy L. and {Thom}, Christopher and {Prochaska}, J. Xavier and {Chen}, Hsiao-Wen},
        title = "{The Last Eight-Billion Years of Intergalactic C IV Evolution}",
      journal = {\apj},
         year = 2010,
        month = jan,
       volume = {708},
       number = {1},
        pages = {868-908},
          doi = {10.1088/0004-637X/708/1/868},
archivePrefix = {arXiv},
       eprint = {0906.3347},
 primaryClass = {astro-ph.CO},
       adsurl = {https://ui.adsabs.harvard.edu/abs/2010ApJ...708..868C}
}

@ARTICLE{sec1meaningfulsampleLehner2014,
       author = {{Lehner}, N. and {O'Meara}, J.~M. and {Fox}, A.~J. and {Howk}, J.~C. and {Prochaska}, J.~X. and {Burns}, V. and {Armstrong}, A.~A.},
        title = "{Galactic and Circumgalactic O VI and its Impact on the Cosmological Metal and Baryon Budgets at 2 < z <\raisebox{-0.5ex}\textasciitilde 3.5}",
      journal = {\apj},
         year = 2014,
        month = jun,
       volume = {788},
       number = {2},
          eid = {119},
        pages = {119},
          doi = {10.1088/0004-637X/788/2/119},
archivePrefix = {arXiv},
       eprint = {1401.1811},
 primaryClass = {astro-ph.CO},
       adsurl = {https://ui.adsabs.harvard.edu/abs/2014ApJ...788..119L}
}

@ARTICLE{sec1meaningfulsampleLiang2014,
       author = {{Liang}, Cameron J. and {Chen}, Hsiao-Wen},
        title = "{Mining circumgalactic baryons in the low-redshift universe}",
      journal = {\mnras},
         year = 2014,
        month = dec,
       volume = {445},
       number = {2},
        pages = {2061-2081},
          doi = {10.1093/mnras/stu1901},
archivePrefix = {arXiv},
       eprint = {1402.3602},
 primaryClass = {astro-ph.CO},
       adsurl = {https://ui.adsabs.harvard.edu/abs/2014MNRAS.445.2061L}
}

@ARTICLE{sec1meaningfulsampleProchaska2011,
       author = {{Prochaska}, J. Xavier and {Weiner}, B. and {Chen}, H.-W. and {Mulchaey}, J. and {Cooksey}, K.},
        title = "{Probing the Intergalactic Medium/Galaxy Connection. V. On the Origin of Ly{\ensuremath{\alpha}} and O VI Absorption at z < 0.2}",
      journal = {\apj},
         year = 2011,
        month = oct,
       volume = {740},
       number = {2},
          eid = {91},
        pages = {91},
          doi = {10.1088/0004-637X/740/2/91},
archivePrefix = {arXiv},
       eprint = {1103.1891},
 primaryClass = {astro-ph.CO},
       adsurl = {https://ui.adsabs.harvard.edu/abs/2011ApJ...740...91P}
}

@ARTICLE{sec1meaningfulsampleTumlinson2013,
       author = {{Tumlinson}, Jason and {Thom}, Christopher and {Werk}, Jessica K. and {Prochaska}, J. Xavier and {Tripp}, Todd M. and {Katz}, Neal and {Dav{\'e}}, Romeel and {Oppenheimer}, Benjamin D. and {Meiring}, Joseph D. and {Ford}, Amanda Brady and {O'Meara}, John M. and {Peeples}, Molly S. and {Sembach}, Kenneth R. and {Weinberg}, David H.},
        title = "{The COS-Halos Survey: Rationale, Design, and a Census of Circumgalactic Neutral Hydrogen}",
      journal = {\apj},
         year = 2013,
        month = nov,
       volume = {777},
       number = {1},
          eid = {59},
        pages = {59},
          doi = {10.1088/0004-637X/777/1/59},
archivePrefix = {arXiv},
       eprint = {1309.6317},
 primaryClass = {astro-ph.CO},
       adsurl = {https://ui.adsabs.harvard.edu/abs/2013ApJ...777...59T}
}

@ARTICLE{sec1meaningfulsampleWerk2016,
       author = {{Werk}, Jessica K. and {Prochaska}, J. Xavier and {Cantalupo}, Sebastiano and {Fox}, Andrew J. and {Oppenheimer}, Benjamin and {Tumlinson}, Jason and {Tripp}, Todd M. and {Lehner}, Nicolas and {McQuinn}, Matthew},
        title = "{The COS-Halos Survey: Origins of the Highly Ionized Circumgalactic Medium of Star-Forming Galaxies}",
      journal = {\apj},
         year = 2016,
        month = dec,
       volume = {833},
       number = {1},
          eid = {54},
        pages = {54},
          doi = {10.3847/1538-4357/833/1/54},
archivePrefix = {arXiv},
       eprint = {1609.00012},
 primaryClass = {astro-ph.GA},
       adsurl = {https://ui.adsabs.harvard.edu/abs/2016ApJ...833...54W}
}

@ARTICLE{sec1meaningfulsampleLehner2020,
       author = {{Lehner}, Nicolas and {Berek}, Samantha C. and {Howk}, J. Christopher and {Wakker}, Bart P. and {Tumlinson}, Jason and {Jenkins}, Edward B. and {Prochaska}, J. Xavier and {Augustin}, Ramona and {Ji}, Suoqing and {Faucher-Gigu{\`e}re}, Claude-Andr{\'e} and {Hafen}, Zachary and {Peeples}, Molly S. and {Barger}, Kat A. and {Berg}, Michelle A. and {Bordoloi}, Rongmon and {Brown}, Thomas M. and {Fox}, Andrew J. and {Gilbert}, Karoline M. and {Guhathakurta}, Puragra and {Kalirai}, Jason S. and {Lockman}, Felix J. and {O'Meara}, John M. and {Pisano}, D.~J. and {Ribaudo}, Joseph and {Werk}, Jessica K.},
        title = "{Project AMIGA: The Circumgalactic Medium of Andromeda}",
      journal = {\apj},
         year = 2020,
        month = sep,
       volume = {900},
       number = {1},
          eid = {9},
        pages = {9},
          doi = {10.3847/1538-4357/aba49c},
archivePrefix = {arXiv},
       eprint = {2002.07818},
 primaryClass = {astro-ph.GA},
       adsurl = {https://ui.adsabs.harvard.edu/abs/2020ApJ...900....9L}
}

@ARTICLE{sec1meaningfulsampleLi2020,
       author = {{Li}, Jiang-Tao},
        title = "{An X{\ensuremath{-}}ray view of the hot circum{\ensuremath{-}}galactic medium}",
      journal = {Astronomische Nachrichten},
         year = 2020,
        month = feb,
       volume = {341},
       number = {2},
        pages = {177-183},
          doi = {10.1002/asna.202023775},
archivePrefix = {arXiv},
       eprint = {2002.08257},
 primaryClass = {astro-ph.HE},
       adsurl = {https://ui.adsabs.harvard.edu/abs/2020AN....341..177L}
}

@ARTICLE{sec1meaningfulsampleZhang2025,
       author = {{Zhang}, Yi and {Comparat}, Johan and {Ponti}, Gabriele and {Merloni}, Andrea and {Nandra}, Kirpal and {Haberl}, Frank and {Truong}, Nhut and {Pillepich}, Annalisa and {Popesso}, Paola and {Locatelli}, Nicola and {Zhang}, Xiaoyuan and {Sanders}, Jeremy and {Zheng}, Xueying and {Liu}, Ang and {Liu}, Teng and {Predehl}, Peter and {Salvato}, Mara and {Bruggen}, Marcus and {Shreeram}, Soumya and {Yeung}, Michael C.~H.},
        title = "{The hot circumgalactic medium in the eROSITA All-Sky Survey: III. Star-forming and quiescent galaxies}",
      journal = {\aap},
         year = 2025,
        month = jan,
       volume = {693},
          eid = {A197},
        pages = {A197},
          doi = {10.1051/0004-6361/202452273},
archivePrefix = {arXiv},
       eprint = {2411.19945},
 primaryClass = {astro-ph.GA},
       adsurl = {https://ui.adsabs.harvard.edu/abs/2025A&A...693A.197Z}
}

@ARTICLE{TNG100,
       author = {{Pillepich}, Annalisa and {Springel}, Volker and {Nelson}, Dylan and {Genel}, Shy and {Naiman}, Jill and {Pakmor}, R{\"u}diger and {Hernquist}, Lars and {Torrey}, Paul and {Vogelsberger}, Mark and {Weinberger}, Rainer and {Marinacci}, Federico},
        title = "{Simulating galaxy formation with the IllustrisTNG model}",
      journal = {\mnras},
         year = 2018,
        month = jan,
       volume = {473},
       number = {3},
        pages = {4077-4106},
          doi = {10.1093/mnras/stx2656},
archivePrefix = {arXiv},
       eprint = {1703.02970},
 primaryClass = {astro-ph.GA},
       adsurl = {https://ui.adsabs.harvard.edu/abs/2018MNRAS.473.4077P}
}

@ARTICLE{TNG50Nelson,
       author = {{Nelson}, Dylan and {Pillepich}, Annalisa and {Springel}, Volker and {Pakmor}, R{\"u}diger and {Weinberger}, Rainer and {Genel}, Shy and {Torrey}, Paul and {Vogelsberger}, Mark and {Marinacci}, Federico and {Hernquist}, Lars},
        title = "{First results from the TNG50 simulation: galactic outflows driven by supernovae and black hole feedback}",
      journal = {\mnras},
         year = 2019,
        month = dec,
       volume = {490},
       number = {3},
        pages = {3234-3261},
          doi = {10.1093/mnras/stz2306},
archivePrefix = {arXiv},
       eprint = {1902.05554},
 primaryClass = {astro-ph.GA},
       adsurl = {https://ui.adsabs.harvard.edu/abs/2019MNRAS.490.3234N}
}

@ARTICLE{TNG50Phillepich,
       author = {{Pillepich}, Annalisa and {Nelson}, Dylan and {Springel}, Volker and {Pakmor}, R{\"u}diger and {Torrey}, Paul and {Weinberger}, Rainer and {Vogelsberger}, Mark and {Marinacci}, Federico and {Genel}, Shy and {van der Wel}, Arjen and {Hernquist}, Lars},
        title = "{First results from the TNG50 simulation: the evolution of stellar and gaseous discs across cosmic time}",
      journal = {\mnras},
         year = 2019,
        month = dec,
       volume = {490},
       number = {3},
        pages = {3196-3233},
          doi = {10.1093/mnras/stz2338},
archivePrefix = {arXiv},
       eprint = {1902.05553},
 primaryClass = {astro-ph.GA},
       adsurl = {https://ui.adsabs.harvard.edu/abs/2019MNRAS.490.3196P}
}

@ARTICLE{EAGLE,
       author = {{Schaye}, Joop and {Crain}, Robert A. and {Bower}, Richard G. and {Furlong}, Michelle and {Schaller}, Matthieu and {Theuns}, Tom and {Dalla Vecchia}, Claudio and {Frenk}, Carlos S. and {McCarthy}, I.~G. and {Helly}, John C. and {Jenkins}, Adrian and {Rosas-Guevara}, Y.~M. and {White}, Simon D.~M. and {Baes}, Maarten and {Booth}, C.~M. and {Camps}, Peter and {Navarro}, Julio F. and {Qu}, Yan and {Rahmati}, Alireza and {Sawala}, Till and {Thomas}, Peter A. and {Trayford}, James},
        title = "{The EAGLE project: simulating the evolution and assembly of galaxies and their environments}",
      journal = {\mnras},
         year = 2015,
        month = jan,
       volume = {446},
       number = {1},
        pages = {521-554},
          doi = {10.1093/mnras/stu2058},
archivePrefix = {arXiv},
       eprint = {1407.7040},
 primaryClass = {astro-ph.GA},
       adsurl = {https://ui.adsabs.harvard.edu/abs/2015MNRAS.446..521S}
}

@ARTICLE{COLIBRE1,
       author = {{Schaye}, Joop and {Chaikin}, Evgenii and {Schaller}, Matthieu and {Ploeckinger}, Sylvia and {Hu{\v{s}}ko}, Filip and {McGibbon}, Rob and {Trayford}, James W. and {Ben{\'\i}tez-Llambay}, Alejandro and {Correa}, Camila and {Frenk}, Carlos S. and {Richings}, Alexander J. and {Forouhar Moreno}, Victor J. and {Bah{\'e}}, Yannick M. and {Borrow}, Josh and {Durrant}, Anna and {Gebek}, Andrea and {Helly}, John C. and {Jenkins}, Adrian and {Lacey}, Cedric G. and {Ludlow}, Aaron and {Nobels}, Folkert S.~J.},
        title = "{The COLIBRE project: cosmological hydrodynamical simulations of galaxy formation and evolution}",
      journal = {arXiv e-prints},
         year = 2025,
        month = aug,
          eid = {arXiv:2508.21126},
        pages = {arXiv:2508.21126},
          doi = {10.48550/arXiv.2508.21126},
archivePrefix = {arXiv},
       eprint = {2508.21126},
 primaryClass = {astro-ph.GA},
       adsurl = {https://ui.adsabs.harvard.edu/abs/2025arXiv250821126S}
}

@ARTICLE{COLIBRE2,
       author = {{Chaikin}, Evgenii and {Schaye}, Joop and {Schaller}, Matthieu and {Ploeckinger}, Sylvia and {Bah{\'e}}, Yannick M. and {Ben{\'\i}tez-Llambay}, Alejandro and {Correa}, Camila and {Forouhar Moreno}, Victor J. and {Frenk}, Carlos S. and {Hu{\v{s}}ko}, Filip and {Kugel}, Roi and {McGibbon}, Robert and {Richings}, Alexander J. and {Trayford}, James W. and {Borrow}, Josh and {Crain}, Robert A. and {Helly}, John C. and {Lacey}, Cedric G. and {Ludlow}, Aaron and {Nobels}, Folkert S.~J.},
        title = "{COLIBRE: calibrating subgrid feedback in cosmological simulations that include a cold gas phase}",
      journal = {arXiv e-prints},
         year = 2025,
        month = sep,
          eid = {arXiv:2509.04067},
        pages = {arXiv:2509.04067},
          doi = {10.48550/arXiv.2509.04067},
archivePrefix = {arXiv},
       eprint = {2509.04067},
 primaryClass = {astro-ph.GA},
       adsurl = {https://ui.adsabs.harvard.edu/abs/2025arXiv250904067C}
}

@ARTICLE{FIRE,
       author = {{Hopkins}, Philip F. and {Wetzel}, Andrew and {Kere{\v{s}}}, Du{\v{s}}an and {Faucher-Gigu{\`e}re}, Claude-Andr{\'e} and {Quataert}, Eliot and {Boylan-Kolchin}, Michael and {Murray}, Norman and {Hayward}, Christopher C. and {Garrison-Kimmel}, Shea and {Hummels}, Cameron and {Feldmann}, Robert and {Torrey}, Paul and {Ma}, Xiangcheng and {Angl{\'e}s-Alc{\'a}zar}, Daniel and {Su}, Kung-Yi and {Orr}, Matthew and {Schmitz}, Denise and {Escala}, Ivanna and {Sanderson}, Robyn and {Grudi{\'c}}, Michael Y. and {Hafen}, Zachary and {Kim}, Ji-Hoon and {Fitts}, Alex and {Bullock}, James S. and {Wheeler}, Coral and {Chan}, T.~K. and {Elbert}, Oliver D. and {Narayanan}, Desika},
        title = "{FIRE-2 simulations: physics versus numerics in galaxy formation}",
      journal = {\mnras},
         year = 2018,
        month = oct,
       volume = {480},
       number = {1},
        pages = {800-863},
          doi = {10.1093/mnras/sty1690},
archivePrefix = {arXiv},
       eprint = {1702.06148},
 primaryClass = {astro-ph.GA},
       adsurl = {https://ui.adsabs.harvard.edu/abs/2018MNRAS.480..800H}
}

@ARTICLE{Auriga,
       author = {{Grand}, Robert J.~J. and {G{\'o}mez}, Facundo A. and {Marinacci}, Federico and {Pakmor}, R{\"u}diger and {Springel}, Volker and {Campbell}, David J.~R. and {Frenk}, Carlos S. and {Jenkins}, Adrian and {White}, Simon D.~M.},
        title = "{The Auriga Project: the properties and formation mechanisms of disc galaxies across cosmic time}",
      journal = {\mnras},
         year = 2017,
        month = may,
       volume = {467},
       number = {1},
        pages = {179-207},
          doi = {10.1093/mnras/stx071},
archivePrefix = {arXiv},
       eprint = {1610.01159},
 primaryClass = {astro-ph.GA},
       adsurl = {https://ui.adsabs.harvard.edu/abs/2017MNRAS.467..179G}
}

@ARTICLE{NIHAO,
       author = {{Wang}, Liang and {Dutton}, Aaron A. and {Stinson}, Gregory S. and {Macci{\`o}}, Andrea V. and {Penzo}, Camilla and {Kang}, Xi and {Keller}, Ben W. and {Wadsley}, James},
        title = "{NIHAO project - I. Reproducing the inefficiency of galaxy formation across cosmic time with a large sample of cosmological hydrodynamical simulations}",
      journal = {\mnras},
         year = 2015,
        month = nov,
       volume = {454},
       number = {1},
        pages = {83-94},
          doi = {10.1093/mnras/stv1937},
archivePrefix = {arXiv},
       eprint = {1503.04818},
 primaryClass = {astro-ph.GA},
       adsurl = {https://ui.adsabs.harvard.edu/abs/2015MNRAS.454...83W}
}

@ARTICLE{sec1idealsimBirnboim2003,
       author = {{Birnboim}, Yuval and {Dekel}, Avishai},
        title = "{Virial shocks in galactic haloes?}",
      journal = {\mnras},
         year = 2003,
        month = oct,
       volume = {345},
       number = {1},
        pages = {349-364},
          doi = {10.1046/j.1365-8711.2003.06955.x},
archivePrefix = {arXiv},
       eprint = {astro-ph/0302161},
 primaryClass = {astro-ph},
       adsurl = {https://ui.adsabs.harvard.edu/abs/2003MNRAS.345..349B}
}

@ARTICLE{sec1idealsimGronke2018,
       author = {{Gronke}, Max and {Oh}, S. Peng},
        title = "{The growth and entrainment of cold gas in a hot wind}",
      journal = {\mnras},
         year = 2018,
        month = oct,
       volume = {480},
       number = {1},
        pages = {L111-L115},
          doi = {10.1093/mnrasl/sly131},
archivePrefix = {arXiv},
       eprint = {1806.02728},
 primaryClass = {astro-ph.GA},
       adsurl = {https://ui.adsabs.harvard.edu/abs/2018MNRAS.480L.111G}
}

@ARTICLE{sec1idealsimFielding2017,
       author = {{Fielding}, Drummond and {Quataert}, Eliot and {McCourt}, Michael and {Thompson}, Todd A.},
        title = "{The impact of star formation feedback on the circumgalactic medium}",
      journal = {\mnras},
         year = 2017,
        month = apr,
       volume = {466},
       number = {4},
        pages = {3810-3826},
          doi = {10.1093/mnras/stw3326},
archivePrefix = {arXiv},
       eprint = {1606.06734},
 primaryClass = {astro-ph.GA},
       adsurl = {https://ui.adsabs.harvard.edu/abs/2017MNRAS.466.3810F}
}

@ARTICLE{sec1idealsimStern2019,
       author = {{Stern}, Jonathan and {Fielding}, Drummond and {Faucher-Gigu{\`e}re}, Claude-Andr{\'e} and {Quataert}, Eliot},
        title = "{Cooling flow solutions for the circumgalactic medium}",
      journal = {\mnras},
         year = 2019,
        month = sep,
       volume = {488},
       number = {2},
        pages = {2549-2572},
          doi = {10.1093/mnras/stz1859},
archivePrefix = {arXiv},
       eprint = {1906.07737},
 primaryClass = {astro-ph.GA},
       adsurl = {https://ui.adsabs.harvard.edu/abs/2019MNRAS.488.2549S}
}

@ARTICLE{sec1idealsimStern2020,
       author = {{Stern}, Jonathan and {Fielding}, Drummond and {Faucher-Gigu{\`e}re}, Claude-Andr{\'e} and {Quataert}, Eliot},
        title = "{The maximum accretion rate of hot gas in dark matter haloes}",
      journal = {\mnras},
         year = 2020,
        month = mar,
       volume = {492},
       number = {4},
        pages = {6042-6058},
          doi = {10.1093/mnras/staa198},
archivePrefix = {arXiv},
       eprint = {1909.07402},
 primaryClass = {astro-ph.GA},
       adsurl = {https://ui.adsabs.harvard.edu/abs/2020MNRAS.492.6042S}
}

@ARTICLE{sec1reproduceobsNelson2018,
       author = {{Nelson}, Dylan and {Kauffmann}, Guinevere and {Pillepich}, Annalisa and {Genel}, Shy and {Springel}, Volker and {Pakmor}, R{\"u}diger and {Hernquist}, Lars and {Weinberger}, Rainer and {Torrey}, Paul and {Vogelsberger}, Mark and {Marinacci}, Federico},
        title = "{The abundance, distribution, and physical nature of highly ionized oxygen O VI, O VII, and O VIII in IllustrisTNG}",
      journal = {\mnras},
         year = 2018,
        month = jun,
       volume = {477},
       number = {1},
        pages = {450-479},
          doi = {10.1093/mnras/sty656},
archivePrefix = {arXiv},
       eprint = {1712.00016},
 primaryClass = {astro-ph.GA},
       adsurl = {https://ui.adsabs.harvard.edu/abs/2018MNRAS.477..450N}
}

@ARTICLE{sec1reproduceobsHafen2017,
       author = {{Hafen}, Zachary and {Faucher-Gigu{\`e}re}, Claude-Andr{\'e} and {Angl{\'e}s-Alc{\'a}zar}, Daniel and {Kere{\v{s}}}, Du{\v{s}}an and {Feldmann}, Robert and {Chan}, T.~K. and {Quataert}, Eliot and {Murray}, Norman and {Hopkins}, Philip F.},
        title = "{Low-redshift Lyman limit systems as diagnostics of cosmological inflows and outflows}",
      journal = {\mnras},
         year = 2017,
        month = aug,
       volume = {469},
       number = {2},
        pages = {2292-2304},
          doi = {10.1093/mnras/stx952},
archivePrefix = {arXiv},
       eprint = {1608.05712},
 primaryClass = {astro-ph.GA},
       adsurl = {https://ui.adsabs.harvard.edu/abs/2017MNRAS.469.2292H}
}

@ARTICLE{sec1reproduceobsGutcke2017,
       author = {{Gutcke}, Thales A. and {Stinson}, Greg S. and {Macci{\`o}}, Andrea V. and {Wang}, Liang and {Dutton}, Aaron A.},
        title = "{NIHAO - VIII. Circum-galactic medium and outflows - The puzzles of H I and O VI gas distributions}",
      journal = {\mnras},
         year = 2017,
        month = jan,
       volume = {464},
       number = {3},
        pages = {2796-2815},
          doi = {10.1093/mnras/stw2539},
archivePrefix = {arXiv},
       eprint = {1602.06956},
 primaryClass = {astro-ph.GA},
       adsurl = {https://ui.adsabs.harvard.edu/abs/2017MNRAS.464.2796G}
}

@ARTICLE{sec1haloshapeLochhaas2020,
       author = {{Lochhaas}, Cassandra and {Bryan}, Greg L. and {Li}, Yuan and {Li}, Miao and {Fielding}, Drummond},
        title = "{Properties of the simulated circumgalactic medium}",
      journal = {\mnras},
         year = 2020,
        month = mar,
       volume = {493},
       number = {1},
        pages = {1461-1478},
          doi = {10.1093/mnras/staa358},
archivePrefix = {arXiv},
       eprint = {1908.00021},
 primaryClass = {astro-ph.GA},
       adsurl = {https://ui.adsabs.harvard.edu/abs/2020MNRAS.493.1461L}
}

@ARTICLE{Dekel2006,
       author = {{Dekel}, Avishai and {Birnboim}, Yuval},
        title = "{Galaxy bimodality due to cold flows and shock heating}",
      journal = {\mnras},
         year = 2006,
        month = may,
       volume = {368},
       number = {1},
        pages = {2-20},
          doi = {10.1111/j.1365-2966.2006.10145.x},
archivePrefix = {arXiv},
       eprint = {astro-ph/0412300},
 primaryClass = {astro-ph},
       adsurl = {https://ui.adsabs.harvard.edu/abs/2006MNRAS.368....2D}
}

@ARTICLE{sec1haloshapeOppenheimer2008,
       author = {{Oppenheimer}, Benjamin D. and {Dav{\'e}}, Romeel},
        title = "{Mass, metal, and energy feedback in cosmological simulations}",
      journal = {\mnras},
         year = 2008,
        month = jun,
       volume = {387},
       number = {2},
        pages = {577-600},
          doi = {10.1111/j.1365-2966.2008.13280.x},
archivePrefix = {arXiv},
       eprint = {0712.1827},
 primaryClass = {astro-ph},
       adsurl = {https://ui.adsabs.harvard.edu/abs/2008MNRAS.387..577O}
}

@ARTICLE{sec1haloshapeOppenheimer2010,
       author = {{Oppenheimer}, Benjamin D. and {Dav{\'e}}, Romeel and {Kere{\v{s}}}, Du{\v{s}}an and {Fardal}, Mark and {Katz}, Neal and {Kollmeier}, Juna A. and {Weinberg}, David H.},
        title = "{Feedback and recycled wind accretion: assembling the z = 0 galaxy mass function}",
      journal = {\mnras},
         year = 2010,
        month = aug,
       volume = {406},
       number = {4},
        pages = {2325-2338},
          doi = {10.1111/j.1365-2966.2010.16872.x},
archivePrefix = {arXiv},
       eprint = {0912.0519},
 primaryClass = {astro-ph.CO},
       adsurl = {https://ui.adsabs.harvard.edu/abs/2010MNRAS.406.2325O}
}

@ARTICLE{sec1haloshapeMuratov2015,
       author = {{Muratov}, Alexander L. and {Kere{\v{s}}}, Du{\v{s}}an and {Faucher-Gigu{\`e}re}, Claude-Andr{\'e} and {Hopkins}, Philip F. and {Quataert}, Eliot and {Murray}, Norman},
        title = "{Gusty, gaseous flows of FIRE: galactic winds in cosmological simulations with explicit stellar feedback}",
      journal = {\mnras},
         year = 2015,
        month = dec,
       volume = {454},
       number = {3},
        pages = {2691-2713},
          doi = {10.1093/mnras/stv2126},
archivePrefix = {arXiv},
       eprint = {1501.03155},
 primaryClass = {astro-ph.GA},
       adsurl = {https://ui.adsabs.harvard.edu/abs/2015MNRAS.454.2691M}
}

@ARTICLE{AA2017,
       author = {{Angl{\'e}s-Alc{\'a}zar}, Daniel and {Faucher-Gigu{\`e}re}, Claude-Andr{\'e} and {Kere{\v{s}}}, Du{\v{s}}an and {Hopkins}, Philip F. and {Quataert}, Eliot and {Murray}, Norman},
        title = "{The cosmic baryon cycle and galaxy mass assembly in the FIRE simulations}",
      journal = {\mnras},
         year = 2017,
        month = oct,
       volume = {470},
       number = {4},
        pages = {4698-4719},
          doi = {10.1093/mnras/stx1517},
archivePrefix = {arXiv},
       eprint = {1610.08523},
 primaryClass = {astro-ph.GA},
       adsurl = {https://ui.adsabs.harvard.edu/abs/2017MNRAS.470.4698A}
}

@ARTICLE{sec1haloshapemetalTumlinson2011,
       author = {{Tumlinson}, J. and {Thom}, C. and {Werk}, J.~K. and {Prochaska}, J.~X. and {Tripp}, T.~M. and {Weinberg}, D.~H. and {Peeples}, M.~S. and {O'Meara}, J.~M. and {Oppenheimer}, B.~D. and {Meiring}, J.~D. and {Katz}, N.~S. and {Dav{\'e}}, R. and {Ford}, A.~B. and {Sembach}, K.~R.},
        title = "{The Large, Oxygen-Rich Halos of Star-Forming Galaxies Are a Major Reservoir of Galactic Metals}",
      journal = {Science},
         year = 2011,
        month = nov,
       volume = {334},
       number = {6058},
        pages = {948},
          doi = {10.1126/science.1209840},
archivePrefix = {arXiv},
       eprint = {1111.3980},
 primaryClass = {astro-ph.CO},
       adsurl = {https://ui.adsabs.harvard.edu/abs/2011Sci...334..948T}
}

@ARTICLE{sec1haloshapemetalProchaska2017,
       author = {{Prochaska}, J. Xavier and {Werk}, Jessica K. and {Worseck}, G{\'a}bor and {Tripp}, Todd M. and {Tumlinson}, Jason and {Burchett}, Joseph N. and {Fox}, Andrew J. and {Fumagalli}, Michele and {Lehner}, Nicolas and {Peeples}, Molly S. and {Tejos}, Nicolas},
        title = "{The COS-Halos Survey: Metallicities in the Low-redshift Circumgalactic Medium}",
      journal = {\apj},
         year = 2017,
        month = mar,
       volume = {837},
       number = {2},
          eid = {169},
        pages = {169},
          doi = {10.3847/1538-4357/aa6007},
archivePrefix = {arXiv},
       eprint = {1702.02618},
 primaryClass = {astro-ph.GA},
       adsurl = {https://ui.adsabs.harvard.edu/abs/2017ApJ...837..169P}
}

@ARTICLE{Hafen2019,
       author = {{Hafen}, Zachary and {Faucher-Gigu{\`e}re}, Claude-Andr{\'e} and {Angl{\'e}s-Alc{\'a}zar}, Daniel and {Stern}, Jonathan and {Kere{\v{s}}}, Du{\v{s}}an and {Hummels}, Cameron and {Esmerian}, Clarke and {Garrison-Kimmel}, Shea and {El-Badry}, Kareem and {Wetzel}, Andrew and {Chan}, T.~K. and {Hopkins}, Philip F. and {Murray}, Norman},
        title = "{The origins of the circumgalactic medium in the FIRE simulations}",
      journal = {\mnras},
         year = 2019,
        month = sep,
       volume = {488},
       number = {1},
        pages = {1248-1272},
          doi = {10.1093/mnras/stz1773},
archivePrefix = {arXiv},
       eprint = {1811.11753},
 primaryClass = {astro-ph.GA},
       adsurl = {https://ui.adsabs.harvard.edu/abs/2019MNRAS.488.1248H}
}

@ARTICLE{Bordoloi2018ApJ...864..132B,
       author = {{Bordoloi}, Rongmon and {Prochaska}, J. Xavier and {Tumlinson}, Jason and {Werk}, Jessica K. and {Tripp}, Todd M. and {Burchett}, Joseph N.},
        title = "{On the CGM Fundamental Plane: The Halo Mass Dependency of Circumgalactic H I}",
      journal = {\apj},
         year = 2018,
        month = sep,
       volume = {864},
       number = {2},
          eid = {132},
        pages = {132},
          doi = {10.3847/1538-4357/aad8ac},
archivePrefix = {arXiv},
       eprint = {1712.02348},
 primaryClass = {astro-ph.GA},
       adsurl = {https://ui.adsabs.harvard.edu/abs/2018ApJ...864..132B}
}

@ARTICLE{Lim2016MNRAS.455..499L,
       author = {{Lim}, S.~H. and {Mo}, H.~J. and {Wang}, Huiyuan and {Yang}, Xiaohu},
        title = "{An observational proxy of halo assembly time and its correlation with galaxy properties}",
      journal = {\mnras},
         year = 2016,
        month = jan,
       volume = {455},
       number = {1},
        pages = {499-510},
          doi = {10.1093/mnras/stv2282},
archivePrefix = {arXiv},
       eprint = {1502.01256},
 primaryClass = {astro-ph.GA},
       adsurl = {https://ui.adsabs.harvard.edu/abs/2016MNRAS.455..499L}
}

@ARTICLE{Tojeiro2017MNRAS.470.3720T,
       author = {{Tojeiro}, Rita and {Eardley}, Elizabeth and {Peacock}, John A. and {Norberg}, Peder and {Alpaslan}, Mehmet and {Driver}, Simon P. and {Henriques}, Bruno and {Hopkins}, Andrew M. and {Kafle}, Prajwal R. and {Robotham}, Aaron S.~G. and {Thomas}, Peter and {Tonini}, Chiara and {Wild}, Vivienne},
        title = "{Galaxy and Mass Assembly (GAMA): halo formation times and halo assembly bias on the cosmic web}",
      journal = {\mnras},
         year = 2017,
        month = sep,
       volume = {470},
       number = {3},
        pages = {3720-3741},
          doi = {10.1093/mnras/stx1466},
archivePrefix = {arXiv},
       eprint = {1612.08595},
 primaryClass = {astro-ph.CO},
       adsurl = {https://ui.adsabs.harvard.edu/abs/2017MNRAS.470.3720T}
}

@ARTICLE{Montero2021MNRAS.508..940M,
       author = {{Montero-Dorta}, Antonio D. and {Chaves-Montero}, Jon{\'a}s and {Artale}, M. Celeste and {Favole}, Ginevra},
        title = "{On the influence of halo mass accretion history on galaxy properties and assembly bias}",
      journal = {\mnras},
         year = 2021,
        month = nov,
       volume = {508},
       number = {1},
        pages = {940-949},
          doi = {10.1093/mnras/stab2556},
archivePrefix = {arXiv},
       eprint = {2105.05274},
 primaryClass = {astro-ph.GA},
       adsurl = {https://ui.adsabs.harvard.edu/abs/2021MNRAS.508..940M}
}

@ARTICLE{sec1smbhDavies2020,
       author = {{Davies}, Jonathan J. and {Crain}, Robert A. and {Oppenheimer}, Benjamin D. and {Schaye}, Joop},
        title = "{The quenching and morphological evolution of central galaxies is facilitated by the feedback-driven expulsion of circumgalactic gas}",
      journal = {\mnras},
         year = 2020,
        month = jan,
       volume = {491},
       number = {3},
        pages = {4462-4480},
          doi = {10.1093/mnras/stz3201},
archivePrefix = {arXiv},
       eprint = {1908.11380},
 primaryClass = {astro-ph.GA},
       adsurl = {https://ui.adsabs.harvard.edu/abs/2020MNRAS.491.4462D}
}

@ARTICLE{sec1smbhTruong2020,
       author = {{Truong}, Nhut and {Pillepich}, Annalisa and {Werner}, Norbert and {Nelson}, Dylan and {Lakhchaura}, Kiran and {Weinberger}, Rainer and {Springel}, Volker and {Vogelsberger}, Mark and {Hernquist}, Lars},
        title = "{X-ray signatures of black hole feedback: hot galactic atmospheres in IllustrisTNG and X-ray observations}",
      journal = {\mnras},
         year = 2020,
        month = may,
       volume = {494},
       number = {1},
        pages = {549-570},
          doi = {10.1093/mnras/staa685},
archivePrefix = {arXiv},
       eprint = {1911.11165},
 primaryClass = {astro-ph.GA},
       adsurl = {https://ui.adsabs.harvard.edu/abs/2020MNRAS.494..549T}
}

@ARTICLE{sec1smbhZinger2020,
       author = {{Zinger}, Elad and {Pillepich}, Annalisa and {Nelson}, Dylan and {Weinberger}, Rainer and {Pakmor}, R{\"u}diger and {Springel}, Volker and {Hernquist}, Lars and {Marinacci}, Federico and {Vogelsberger}, Mark},
        title = "{Ejective and preventative: the IllustrisTNG black hole feedback and its effects on the thermodynamics of the gas within and around galaxies}",
      journal = {\mnras},
         year = 2020,
        month = nov,
       volume = {499},
       number = {1},
        pages = {768-792},
          doi = {10.1093/mnras/staa2607},
archivePrefix = {arXiv},
       eprint = {2004.06132},
 primaryClass = {astro-ph.GA},
       adsurl = {https://ui.adsabs.harvard.edu/abs/2020MNRAS.499..768Z}
}

@ARTICLE{sec1smbhOppenheimer2018,
       author = {{Oppenheimer}, Benjamin D.},
        title = "{Deviations from hydrostatic equilibrium in the circumgalactic medium: spinning hot haloes and accelerating flows}",
      journal = {\mnras},
         year = 2018,
        month = nov,
       volume = {480},
       number = {3},
        pages = {2963-2975},
          doi = {10.1093/mnras/sty1918},
archivePrefix = {arXiv},
       eprint = {1801.00788},
 primaryClass = {astro-ph.GA},
       adsurl = {https://ui.adsabs.harvard.edu/abs/2018MNRAS.480.2963O}
}

@ARTICLE{sec1smbhNelson2015,
       author = {{Nelson}, Dylan and {Genel}, Shy and {Vogelsberger}, Mark and {Springel}, Volker and {Sijacki}, Debora and {Torrey}, Paul and {Hernquist}, Lars},
        title = "{The impact of feedback on cosmological gas accretion}",
      journal = {\mnras},
         year = 2015,
        month = mar,
       volume = {448},
       number = {1},
        pages = {59-74},
          doi = {10.1093/mnras/stv017},
archivePrefix = {arXiv},
       eprint = {1410.5425},
 primaryClass = {astro-ph.CO},
       adsurl = {https://ui.adsabs.harvard.edu/abs/2015MNRAS.448...59N}
}

@ARTICLE{sec1envshapeKeres2005,
       author = {{Kere{\v{s}}}, Du{\v{s}}an and {Katz}, Neal and {Weinberg}, David H. and {Dav{\'e}}, Romeel},
        title = "{How do galaxies get their gas?}",
      journal = {\mnras},
         year = 2005,
        month = oct,
       volume = {363},
       number = {1},
        pages = {2-28},
          doi = {10.1111/j.1365-2966.2005.09451.x},
archivePrefix = {arXiv},
       eprint = {astro-ph/0407095},
 primaryClass = {astro-ph},
       adsurl = {https://ui.adsabs.harvard.edu/abs/2005MNRAS.363....2K}
}

@ARTICLE{D19,
       author = {{Davies}, Jonathan J. and {Crain}, Robert A. and {McCarthy}, Ian G. and {Oppenheimer}, Benjamin D. and {Schaye}, Joop and {Schaller}, Matthieu and {McAlpine}, Stuart},
        title = "{The gas fractions of dark matter haloes hosting simulated {\ensuremath{\sim}}L$^{{\ensuremath{\star}}}$ galaxies are governed by the feedback history of their black holes}",
      journal = {\mnras},
         year = 2019,
        month = may,
       volume = {485},
       number = {3},
        pages = {3783-3793},
          doi = {10.1093/mnras/stz635},
archivePrefix = {arXiv},
       eprint = {1810.07696},
 primaryClass = {astro-ph.GA},
       adsurl = {https://ui.adsabs.harvard.edu/abs/2019MNRAS.485.3783D}
}

@ARTICLE{D21,
       author = {{Davies}, Jonathan J. and {Crain}, Robert A. and {Pontzen}, Andrew},
        title = "{Quenching and morphological evolution due to circumgalactic gas expulsion in a simulated galaxy with a controlled assembly history}",
      journal = {\mnras},
         year = 2021,
        month = jan,
       volume = {501},
       number = {1},
        pages = {236-253},
          doi = {10.1093/mnras/staa3643},
archivePrefix = {arXiv},
       eprint = {2006.13221},
 primaryClass = {astro-ph.GA},
       adsurl = {https://ui.adsabs.harvard.edu/abs/2021MNRAS.501..236D}
}

@ARTICLE{Gao2005,
       author = {{Gao}, Liang and {Springel}, Volker and {White}, Simon D.~M.},
        title = "{The age dependence of halo clustering}",
      journal = {\mnras},
         year = 2005,
        month = oct,
       volume = {363},
       number = {1},
        pages = {L66-L70},
          doi = {10.1111/j.1745-3933.2005.00084.x},
archivePrefix = {arXiv},
       eprint = {astro-ph/0506510},
 primaryClass = {astro-ph},
       adsurl = {https://ui.adsabs.harvard.edu/abs/2005MNRAS.363L..66G}
}

@ARTICLE{TNGdata,
       author = {{Nelson}, D. and {Pillepich}, A. and {Genel}, S. and {Vogelsberger}, M. and {Springel}, V. and {Torrey}, P. and {Rodriguez-Gomez}, V. and {Sijacki}, D. and {Snyder}, G.~F. and {Griffen}, B. and {Marinacci}, F. and {Blecha}, L. and {Sales}, L. and {Xu}, D. and {Hernquist}, L.},
        title = "{The illustris simulation: Public data release}",
      journal = {Astronomy and Computing},
         year = 2015,
        month = nov,
       volume = {13},
        pages = {12-37},
          doi = {10.1016/j.ascom.2015.09.003},
archivePrefix = {arXiv},
       eprint = {1504.00362},
 primaryClass = {astro-ph.CO},
       adsurl = {https://ui.adsabs.harvard.edu/abs/2015A&C....13...12N}
}

@ARTICLE{TNG1,
       author = {{Pillepich}, Annalisa and {Nelson}, Dylan and {Hernquist}, Lars and {Springel}, Volker and {Pakmor}, R{\"u}diger and {Torrey}, Paul and {Weinberger}, Rainer and {Genel}, Shy and {Naiman}, Jill P. and {Marinacci}, Federico and {Vogelsberger}, Mark},
        title = "{First results from the IllustrisTNG simulations: the stellar mass content of groups and clusters of galaxies}",
      journal = {\mnras},
         year = 2018,
        month = mar,
       volume = {475},
       number = {1},
        pages = {648-675},
          doi = {10.1093/mnras/stx3112},
archivePrefix = {arXiv},
       eprint = {1707.03406},
 primaryClass = {astro-ph.GA},
       adsurl = {https://ui.adsabs.harvard.edu/abs/2018MNRAS.475..648P}
}

@ARTICLE{TNG2,
       author = {{Springel}, Volker and {Pakmor}, R{\"u}diger and {Pillepich}, Annalisa and {Weinberger}, Rainer and {Nelson}, Dylan and {Hernquist}, Lars and {Vogelsberger}, Mark and {Genel}, Shy and {Torrey}, Paul and {Marinacci}, Federico and {Naiman}, Jill},
        title = "{First results from the IllustrisTNG simulations: matter and galaxy clustering}",
      journal = {\mnras},
         year = 2018,
        month = mar,
       volume = {475},
       number = {1},
        pages = {676-698},
          doi = {10.1093/mnras/stx3304},
archivePrefix = {arXiv},
       eprint = {1707.03397},
 primaryClass = {astro-ph.GA},
       adsurl = {https://ui.adsabs.harvard.edu/abs/2018MNRAS.475..676S}
}

@ARTICLE{TNG3,
       author = {{Marinacci}, Federico and {Vogelsberger}, Mark and {Pakmor}, R{\"u}diger and {Torrey}, Paul and {Springel}, Volker and {Hernquist}, Lars and {Nelson}, Dylan and {Weinberger}, Rainer and {Pillepich}, Annalisa and {Naiman}, Jill and {Genel}, Shy},
        title = "{First results from the IllustrisTNG simulations: radio haloes and magnetic fields}",
      journal = {\mnras},
         year = 2018,
        month = nov,
       volume = {480},
       number = {4},
        pages = {5113-5139},
          doi = {10.1093/mnras/sty2206},
archivePrefix = {arXiv},
       eprint = {1707.03396},
 primaryClass = {astro-ph.CO},
       adsurl = {https://ui.adsabs.harvard.edu/abs/2018MNRAS.480.5113M}
}

@ARTICLE{TNG4,
       author = {{Naiman}, Jill P. and {Pillepich}, Annalisa and {Springel}, Volker and {Ramirez-Ruiz}, Enrico and {Torrey}, Paul and {Vogelsberger}, Mark and {Pakmor}, R{\"u}diger and {Nelson}, Dylan and {Marinacci}, Federico and {Hernquist}, Lars and {Weinberger}, Rainer and {Genel}, Shy},
        title = "{First results from the IllustrisTNG simulations: a tale of two elements - chemical evolution of magnesium and europium}",
      journal = {\mnras},
         year = 2018,
        month = jun,
       volume = {477},
       number = {1},
        pages = {1206-1224},
          doi = {10.1093/mnras/sty618},
archivePrefix = {arXiv},
       eprint = {1707.03401},
 primaryClass = {astro-ph.GA},
       adsurl = {https://ui.adsabs.harvard.edu/abs/2018MNRAS.477.1206N}
}

@ARTICLE{TNG5,
       author = {{Nelson}, Dylan and {Pillepich}, Annalisa and {Springel}, Volker and {Weinberger}, Rainer and {Hernquist}, Lars and {Pakmor}, R{\"u}diger and {Genel}, Shy and {Torrey}, Paul and {Vogelsberger}, Mark and {Kauffmann}, Guinevere and {Marinacci}, Federico and {Naiman}, Jill},
        title = "{First results from the IllustrisTNG simulations: the galaxy colour bimodality}",
      journal = {\mnras},
         year = 2018,
        month = mar,
       volume = {475},
       number = {1},
        pages = {624-647},
          doi = {10.1093/mnras/stx3040},
archivePrefix = {arXiv},
       eprint = {1707.03395},
 primaryClass = {astro-ph.GA},
       adsurl = {https://ui.adsabs.harvard.edu/abs/2018MNRAS.475..624N}
}

@ARTICLE{arepo2010,
       author = {{Springel}, Volker},
        title = "{E pur si muove: Galilean-invariant cosmological hydrodynamical simulations on a moving mesh}",
      journal = {\mnras},
         year = 2010,
        month = jan,
       volume = {401},
       number = {2},
        pages = {791-851},
          doi = {10.1111/j.1365-2966.2009.15715.x},
archivePrefix = {arXiv},
       eprint = {0901.4107},
 primaryClass = {astro-ph.CO},
       adsurl = {https://ui.adsabs.harvard.edu/abs/2010MNRAS.401..791S}
}

@ARTICLE{subgridmodelbh,
       author = {{Weinberger}, Rainer and {Springel}, Volker and {Hernquist}, Lars and {Pillepich}, Annalisa and {Marinacci}, Federico and {Pakmor}, R{\"u}diger and {Nelson}, Dylan and {Genel}, Shy and {Vogelsberger}, Mark and {Naiman}, Jill and {Torrey}, Paul},
        title = "{Simulating galaxy formation with black hole driven thermal and kinetic feedback}",
      journal = {\mnras},
         year = 2017,
        month = mar,
       volume = {465},
       number = {3},
        pages = {3291-3308},
          doi = {10.1093/mnras/stw2944},
archivePrefix = {arXiv},
       eprint = {1607.03486},
 primaryClass = {astro-ph.GA},
       adsurl = {https://ui.adsabs.harvard.edu/abs/2017MNRAS.465.3291W}
}

@ARTICLE{subgridmodel,
       author = {{Vogelsberger}, Mark and {Genel}, Shy and {Sijacki}, Debora and {Torrey}, Paul and {Springel}, Volker and {Hernquist}, Lars},
        title = "{A model for cosmological simulations of galaxy formation physics}",
      journal = {\mnras},
         year = 2013,
        month = dec,
       volume = {436},
       number = {4},
        pages = {3031-3067},
          doi = {10.1093/mnras/stt1789},
archivePrefix = {arXiv},
       eprint = {1305.2913},
 primaryClass = {astro-ph.CO},
       adsurl = {https://ui.adsabs.harvard.edu/abs/2013MNRAS.436.3031V}
}

@ARTICLE{tracermethod2013,
       author = {{Genel}, Shy and {Vogelsberger}, Mark and {Nelson}, Dylan and {Sijacki}, Debora and {Springel}, Volker and {Hernquist}, Lars},
        title = "{Following the flow: tracer particles in astrophysical fluid simulations}",
      journal = {\mnras},
         year = 2013,
        month = oct,
       volume = {435},
       number = {2},
        pages = {1426-1442},
          doi = {10.1093/mnras/stt1383},
archivePrefix = {arXiv},
       eprint = {1305.2195},
 primaryClass = {astro-ph.IM},
       adsurl = {https://ui.adsabs.harvard.edu/abs/2013MNRAS.435.1426G}
}

@ARTICLE{tracermethod2013Nelson,
       author = {{Nelson}, Dylan and {Vogelsberger}, Mark and {Genel}, Shy and {Sijacki}, Debora and {Kere{\v{s}}}, Du{\v{s}}an and {Springel}, Volker and {Hernquist}, Lars},
        title = "{Moving mesh cosmology: tracing cosmological gas accretion}",
      journal = {\mnras},
         year = 2013,
        month = mar,
       volume = {429},
       number = {4},
        pages = {3353-3370},
          doi = {10.1093/mnras/sts595},
archivePrefix = {arXiv},
       eprint = {1301.6753},
 primaryClass = {astro-ph.CO},
       adsurl = {https://ui.adsabs.harvard.edu/abs/2013MNRAS.429.3353N}
}

@ARTICLE{traceruse2020,
       author = {{Nelson}, Dylan and {Sharma}, Prateek and {Pillepich}, Annalisa and {Springel}, Volker and {Pakmor}, R{\"u}diger and {Weinberger}, Rainer and {Vogelsberger}, Mark and {Marinacci}, Federico and {Hernquist}, Lars},
        title = "{Resolving small-scale cold circumgalactic gas in TNG50}",
      journal = {\mnras},
         year = 2020,
        month = oct,
       volume = {498},
       number = {2},
        pages = {2391-2414},
          doi = {10.1093/mnras/staa2419},
archivePrefix = {arXiv},
       eprint = {2005.09654},
 primaryClass = {astro-ph.GA},
       adsurl = {https://ui.adsabs.harvard.edu/abs/2020MNRAS.498.2391N}
}

@ARTICLE{shotnoise2020,
       author = {{Mitchell}, Peter D. and {Schaye}, Joop and {Bower}, Richard G. and {Crain}, Robert A.},
        title = "{Galactic outflow rates in the EAGLE simulations}",
      journal = {\mnras},
         year = 2020,
        month = may,
       volume = {494},
       number = {3},
        pages = {3971-3997},
          doi = {10.1093/mnras/staa938},
archivePrefix = {arXiv},
       eprint = {1910.09566},
 primaryClass = {astro-ph.GA},
       adsurl = {https://ui.adsabs.harvard.edu/abs/2020MNRAS.494.3971M}
}

@ARTICLE{sec33gasonlycannotFord2016,
       author = {{Ford}, Amanda Brady and {Werk}, Jessica K. and {Dav{\'e}}, Romeel and {Tumlinson}, Jason and {Bordoloi}, Rongmon and {Katz}, Neal and {Kollmeier}, Juna A. and {Oppenheimer}, Benjamin D. and {Peeples}, Molly S. and {Prochaska}, Jason X. and {Weinberg}, David H.},
        title = "{Baryon cycling in the low-redshift circumgalactic medium: a comparison of simulations to the COS-Halos survey}",
      journal = {\mnras},
         year = 2016,
        month = jun,
       volume = {459},
       number = {2},
        pages = {1745-1763},
          doi = {10.1093/mnras/stw595},
archivePrefix = {arXiv},
       eprint = {1503.02084},
 primaryClass = {astro-ph.GA},
       adsurl = {https://ui.adsabs.harvard.edu/abs/2016MNRAS.459.1745F}
}

@ARTICLE{sec33gasonlycannotLehner2013,
       author = {{Lehner}, N. and {Howk}, J.~C. and {Tripp}, T.~M. and {Tumlinson}, J. and {Prochaska}, J.~X. and {O'Meara}, J.~M. and {Thom}, C. and {Werk}, J.~K. and {Fox}, A.~J. and {Ribaudo}, J.},
        title = "{The Bimodal Metallicity Distribution of the Cool Circumgalactic Medium at z <\raisebox{-0.5ex}\textasciitilde 1}",
      journal = {\apj},
         year = 2013,
        month = jun,
       volume = {770},
       number = {2},
          eid = {138},
        pages = {138},
          doi = {10.1088/0004-637X/770/2/138},
archivePrefix = {arXiv},
       eprint = {1302.5424},
 primaryClass = {astro-ph.CO},
       adsurl = {https://ui.adsabs.harvard.edu/abs/2013ApJ...770..138L}
}

@ARTICLE{sec33gasonlycannotWerk2014,
       author = {{Werk}, Jessica K. and {Prochaska}, J. Xavier and {Tumlinson}, Jason and {Peeples}, Molly S. and {Tripp}, Todd M. and {Fox}, Andrew J. and {Lehner}, Nicolas and {Thom}, Christopher and {O'Meara}, John M. and {Ford}, Amanda Brady and {Bordoloi}, Rongmon and {Katz}, Neal and {Tejos}, Nicolas and {Oppenheimer}, Benjamin D. and {Dav{\'e}}, Romeel and {Weinberg}, David H.},
        title = "{The COS-Halos Survey: Physical Conditions and Baryonic Mass in the Low-redshift Circumgalactic Medium}",
      journal = {\apj},
         year = 2014,
        month = sep,
       volume = {792},
       number = {1},
          eid = {8},
        pages = {8},
          doi = {10.1088/0004-637X/792/1/8},
archivePrefix = {arXiv},
       eprint = {1403.0947},
 primaryClass = {astro-ph.CO},
       adsurl = {https://ui.adsabs.harvard.edu/abs/2014ApJ...792....8W}
}

@ARTICLE{sec33metaltracePeeples2014,
       author = {{Peeples}, Molly S. and {Werk}, Jessica K. and {Tumlinson}, Jason and {Oppenheimer}, Benjamin D. and {Prochaska}, J. Xavier and {Katz}, Neal and {Weinberg}, David H.},
        title = "{A Budget and Accounting of Metals at z \raisebox{-0.5ex}\textasciitilde 0: Results from the COS-Halos Survey}",
      journal = {\apj},
         year = 2014,
        month = may,
       volume = {786},
       number = {1},
          eid = {54},
        pages = {54},
          doi = {10.1088/0004-637X/786/1/54},
archivePrefix = {arXiv},
       eprint = {1310.2253},
 primaryClass = {astro-ph.CO},
       adsurl = {https://ui.adsabs.harvard.edu/abs/2014ApJ...786...54P}
}

@ARTICLE{sec33coolingSutherland1993,
       author = {{Sutherland}, Ralph S. and {Dopita}, M.~A.},
        title = "{Cooling Functions for Low-Density Astrophysical Plasmas}",
      journal = {\apjs},
         year = 1993,
        month = sep,
       volume = {88},
        pages = {253},
          doi = {10.1086/191823},
       adsurl = {https://ui.adsabs.harvard.edu/abs/1993ApJS...88..253S}
}

@ARTICLE{sec33coolingWiersma2009,
       author = {{Wiersma}, Robert P.~C. and {Schaye}, Joop and {Smith}, Britton D.},
        title = "{The effect of photoionization on the cooling rates of enriched, astrophysical plasmas}",
      journal = {\mnras},
         year = 2009,
        month = feb,
       volume = {393},
       number = {1},
        pages = {99-107},
          doi = {10.1111/j.1365-2966.2008.14191.x},
archivePrefix = {arXiv},
       eprint = {0807.3748},
 primaryClass = {astro-ph},
       adsurl = {https://ui.adsabs.harvard.edu/abs/2009MNRAS.393...99W}
}

@ARTICLE{sev34Zjupa2017,
       author = {{Zjupa}, Jolanta and {Springel}, Volker},
        title = "{Angular momentum properties of haloes and their baryon content in the Illustris simulation}",
      journal = {\mnras},
         year = 2017,
        month = apr,
       volume = {466},
       number = {2},
        pages = {1625-1647},
          doi = {10.1093/mnras/stw2945},
archivePrefix = {arXiv},
       eprint = {1608.01323},
 primaryClass = {astro-ph.CO},
       adsurl = {https://ui.adsabs.harvard.edu/abs/2017MNRAS.466.1625Z}
}

@ARTICLE{spinparameterBullock2001,
       author = {{Bullock}, J.~S. and {Dekel}, A. and {Kolatt}, T.~S. and {Kravtsov}, A.~V. and {Klypin}, A.~A. and {Porciani}, C. and {Primack}, J.~R.},
        title = "{A Universal Angular Momentum Profile for Galactic Halos}",
      journal = {\apj},
         year = 2001,
        month = jul,
       volume = {555},
       number = {1},
        pages = {240-257},
          doi = {10.1086/321477},
archivePrefix = {arXiv},
       eprint = {astro-ph/0011001},
 primaryClass = {astro-ph},
       adsurl = {https://ui.adsabs.harvard.edu/abs/2001ApJ...555..240B}
}

@ARTICLE{Hahn2009MNRAS.398.1742H,
       author = {{Hahn}, Oliver and {Porciani}, Cristiano and {Dekel}, Avishai and {Carollo}, C. Marcella},
        title = "{Tidal effects and the environment dependence of halo assembly}",
      journal = {\mnras},
         year = 2009,
        month = oct,
       volume = {398},
       number = {4},
        pages = {1742-1756},
          doi = {10.1111/j.1365-2966.2009.15271.x},
archivePrefix = {arXiv},
       eprint = {0803.4211},
 primaryClass = {astro-ph},
       adsurl = {https://ui.adsabs.harvard.edu/abs/2009MNRAS.398.1742H}
}

@ARTICLE{Voort2017MNRAS.466.3460V,
       author = {{van de Voort}, Freeke and {Bah{\'e}}, Yannick M. and {Bower}, Richard G. and {Correa}, Camila A. and {Crain}, Robert A. and {Schaye}, Joop and {Theuns}, Tom},
        title = "{The environmental dependence of gas accretion on to galaxies: quenching satellites through starvation}",
      journal = {\mnras},
         year = 2017,
        month = apr,
       volume = {466},
       number = {3},
        pages = {3460-3471},
          doi = {10.1093/mnras/stw3356},
archivePrefix = {arXiv},
       eprint = {1611.03870},
 primaryClass = {astro-ph.GA},
       adsurl = {https://ui.adsabs.harvard.edu/abs/2017MNRAS.466.3460V}
}

@ARTICLE{Borzyszkowski2017MNRAS.469..594B,
       author = {{Borzyszkowski}, Mikolaj and {Porciani}, Cristiano and {Romano-D{\'\i}az}, Emilio and {Garaldi}, Enrico},
        title = "{ZOMG - I. How the cosmic web inhibits halo growth and generates assembly bias}",
      journal = {\mnras},
         year = 2017,
        month = jul,
       volume = {469},
       number = {1},
        pages = {594-611},
          doi = {10.1093/mnras/stx873},
archivePrefix = {arXiv},
       eprint = {1610.04231},
 primaryClass = {astro-ph.CO},
       adsurl = {https://ui.adsabs.harvard.edu/abs/2017MNRAS.469..594B}
}

@ARTICLE{Fujita2004PASJ...56...29F,
       author = {{Fujita}, Yutaka},
        title = "{Pre-Processing of Galaxies before Entering a Cluster}",
      journal = {\pasj},
         year = 2004,
        month = feb,
       volume = {56},
        pages = {29-43},
          doi = {10.1093/pasj/56.1.29},
archivePrefix = {arXiv},
       eprint = {astro-ph/0311193},
 primaryClass = {astro-ph},
       adsurl = {https://ui.adsabs.harvard.edu/abs/2004PASJ...56...29F}
}

@ARTICLE{Vijayaraghavan2013MNRAS.435.2713V,
       author = {{Vijayaraghavan}, R. and {Ricker}, P.~M.},
        title = "{Pre-processing and post-processing in group-cluster mergers}",
      journal = {\mnras},
         year = 2013,
        month = nov,
       volume = {435},
       number = {3},
        pages = {2713-2735},
          doi = {10.1093/mnras/stt1485},
archivePrefix = {arXiv},
       eprint = {1308.1311},
 primaryClass = {astro-ph.CO},
       adsurl = {https://ui.adsabs.harvard.edu/abs/2013MNRAS.435.2713V}
}

@ARTICLE{Zehavi2018ApJ...853...84Z,
       author = {{Zehavi}, Idit and {Contreras}, Sergio and {Padilla}, Nelson and {Smith}, Nicholas J. and {Baugh}, Carlton M. and {Norberg}, Peder},
        title = "{The Impact of Assembly Bias on the Galaxy Content of Dark Matter Halos}",
      journal = {\apj},
         year = 2018,
        month = jan,
       volume = {853},
       number = {1},
          eid = {84},
        pages = {84},
          doi = {10.3847/1538-4357/aaa54a},
archivePrefix = {arXiv},
       eprint = {1706.07871},
 primaryClass = {astro-ph.GA},
       adsurl = {https://ui.adsabs.harvard.edu/abs/2018ApJ...853...84Z}
}

@ARTICLE{Wechsler2002ApJ...568...52W,
       author = {{Wechsler}, Risa H. and {Bullock}, James S. and {Primack}, Joel R. and {Kravtsov}, Andrey V. and {Dekel}, Avishai},
        title = "{Concentrations of Dark Halos from Their Assembly Histories}",
      journal = {\apj},
         year = 2002,
        month = mar,
       volume = {568},
       number = {1},
        pages = {52-70},
          doi = {10.1086/338765},
archivePrefix = {arXiv},
       eprint = {astro-ph/0108151},
 primaryClass = {astro-ph},
       adsurl = {https://ui.adsabs.harvard.edu/abs/2002ApJ...568...52W}
}

@ARTICLE{eROSITA2021A&A...647A...1P,
       author = {{Predehl}, P. and {Andritschke}, R. and {Arefiev}, V. and {Babyshkin}, V. and {Batanov}, O. and {Becker}, W. and {B{\"o}hringer}, H. and {Bogomolov}, A. and {Boller}, T. and {Borm}, K. and {Bornemann}, W. and {Br{\"a}uninger}, H. and {Br{\"u}ggen}, M. and {Brunner}, H. and {Brusa}, M. and {Bulbul}, E. and {Buntov}, M. and {Burwitz}, V. and {Burkert}, W. and {Clerc}, N. and {Churazov}, E. and {Coutinho}, D. and {Dauser}, T. and {Dennerl}, K. and {Doroshenko}, V. and {Eder}, J. and {Emberger}, V. and {Eraerds}, T. and {Finoguenov}, A. and {Freyberg}, M. and {Friedrich}, P. and {Friedrich}, S. and {F{\"u}rmetz}, M. and {Georgakakis}, A. and {Gilfanov}, M. and {Granato}, S. and {Grossberger}, C. and {Gueguen}, A. and {Gureev}, P. and {Haberl}, F. and {H{\"a}lker}, O. and {Hartner}, G. and {Hasinger}, G. and {Huber}, H. and {Ji}, L. and {Kienlin}, A. v. and {Kink}, W. and {Korotkov}, F. and {Kreykenbohm}, I. and {Lamer}, G. and {Lomakin}, I. and {Lapshov}, I. and {Liu}, T. and {Maitra}, C. and {Meidinger}, N. and {Menz}, B. and {Merloni}, A. and {Mernik}, T. and {Mican}, B. and {Mohr}, J. and {M{\"u}ller}, S. and {Nandra}, K. and {Nazarov}, V. and {Pacaud}, F. and {Pavlinsky}, M. and {Perinati}, E. and {Pfeffermann}, E. and {Pietschner}, D. and {Ramos-Ceja}, M.~E. and {Rau}, A. and {Reiffers}, J. and {Reiprich}, T.~H. and {Robrade}, J. and {Salvato}, M. and {Sanders}, J. and {Santangelo}, A. and {Sasaki}, M. and {Scheuerle}, H. and {Schmid}, C. and {Schmitt}, J. and {Schwope}, A. and {Shirshakov}, A. and {Steinmetz}, M. and {Stewart}, I. and {Str{\"u}der}, L. and {Sunyaev}, R. and {Tenzer}, C. and {Tiedemann}, L. and {Tr{\"u}mper}, J. and {Voron}, V. and {Weber}, P. and {Wilms}, J. and {Yaroshenko}, V.},
        title = "{The eROSITA X-ray telescope on SRG}",
      journal = {\aap},
         year = 2021,
        month = mar,
       volume = {647},
          eid = {A1},
        pages = {A1},
          doi = {10.1051/0004-6361/202039313},
archivePrefix = {arXiv},
       eprint = {2010.03477},
 primaryClass = {astro-ph.HE},
       adsurl = {https://ui.adsabs.harvard.edu/abs/2021A&A...647A...1P}
}

@ARTICLE{Sunayama2016MNRAS.458.1510S,
       author = {{Sunayama}, Tomomi and {Hearin}, Andrew P. and {Padmanabhan}, Nikhil and {Leauthaud}, Alexie},
        title = "{The scale-dependence of halo assembly bias}",
      journal = {\mnras},
         year = 2016,
        month = may,
       volume = {458},
       number = {2},
        pages = {1510-1516},
          doi = {10.1093/mnras/stw332},
archivePrefix = {arXiv},
       eprint = {1509.06417},
 primaryClass = {astro-ph.CO},
       adsurl = {https://ui.adsabs.harvard.edu/abs/2016MNRAS.458.1510S}
}

@ARTICLE{Ramesh2023MNRAS.518.5754R,
       author = {{Ramesh}, Rahul and {Nelson}, Dylan and {Pillepich}, Annalisa},
        title = "{The circumgalactic medium of Milky Way-like galaxies in the TNG50 simulation - I: halo gas properties and the role of SMBH feedback}",
      journal = {\mnras},
         year = 2023,
        month = jan,
       volume = {518},
       number = {4},
        pages = {5754-5777},
          doi = {10.1093/mnras/stac3524},
archivePrefix = {arXiv},
       eprint = {2211.00020},
 primaryClass = {astro-ph.GA},
       adsurl = {https://ui.adsabs.harvard.edu/abs/2023MNRAS.518.5754R}
}

@ARTICLE{Ramesh2023MNRAS.522.1535R,
       author = {{Ramesh}, Rahul and {Nelson}, Dylan and {Pillepich}, Annalisa},
        title = "{The circumgalactic medium of Milky Way-like galaxies in the TNG50 simulation - II. Cold, dense gas clouds and high-velocity cloud analogs}",
      journal = {\mnras},
         year = 2023,
        month = jun,
       volume = {522},
       number = {1},
        pages = {1535-1555},
          doi = {10.1093/mnras/stad951},
archivePrefix = {arXiv},
       eprint = {2303.16215},
 primaryClass = {astro-ph.GA},
       adsurl = {https://ui.adsabs.harvard.edu/abs/2023MNRAS.522.1535R}
}

@ARTICLE{Wang2022MNRAS.509.3148W,
       author = {{Wang}, Sen and {Xu}, Dandan and {Lu}, Shengdong and {Cai}, Zheng and {Xiang}, Maosheng and {Mao}, Shude and {Springel}, Volker and {Hernquist}, Lars},
        title = "{From large-scale environment to CGM angular momentum to star-forming activities - I. Star-forming galaxies}",
      journal = {\mnras},
         year = 2022,
        month = jan,
       volume = {509},
       number = {3},
        pages = {3148-3162},
          doi = {10.1093/mnras/stab3167},
archivePrefix = {arXiv},
       eprint = {2109.06200},
 primaryClass = {astro-ph.GA},
       adsurl = {https://ui.adsabs.harvard.edu/abs/2022MNRAS.509.3148W}
}

@ARTICLE{Lu2022MNRAS.509.2707L,
       author = {{Lu}, Shengdong and {Xu}, Dandan and {Wang}, Sen and {Cai}, Zheng and {He}, Chuan and {Xu}, C. Kevin and {Xia}, Xiaoyang and {Mao}, Shude and {Springel}, Volker and {Hernquist}, Lars},
        title = "{From large-scale environment to CGM angular momentum to star forming activities - II. Quenched galaxies}",
      journal = {\mnras},
         year = 2022,
        month = jan,
       volume = {509},
       number = {2},
        pages = {2707-2719},
          doi = {10.1093/mnras/stab3169},
archivePrefix = {arXiv},
       eprint = {2109.06197},
 primaryClass = {astro-ph.GA},
       adsurl = {https://ui.adsabs.harvard.edu/abs/2022MNRAS.509.2707L}
}
\bibliographystyle{aasjournalv7}

\appendix
\section{Permutation Test}
\label{appendix:perm}

Permutation test is a classical non-parametric statistical tool and has also been used in astronomical researches to assess whether two samples are statistically distinguishable \citep{Morell2020MNRAS.494.3317M,Alpaslan2021MNRAS.505.5403A,deSa2022MNRAS.509.3889D}. Specifically, we define a separation statistic
\begin{equation}
T=\frac{1}{N_r}\sum_{i=1}^{N_r}\left|\log_{10}\tilde{y}_{\rm late}(r_i)-\log_{10}\tilde{y}_{\rm early}(r_i)\right|,
\end{equation}
where $r_i$ denotes the $i$th radial bin, $N_r$ is the total number of radial bins, and $\tilde{y}$ represents the median profile obtained from a   realization of the corresponding samples. We first estimate the observed separation between the `early' and `late' median trends by bootstrap resampling the original samples 2000 times, from which we obtain a distribution of $T$ values, denoted $T_{\rm sim}$. We then construct a corresponding null distribution by randomly shuffling the `early' and `late' labels, while keeping the sample sizes fixed, and repeating the same bootstrap procedure 2000 times to obtain $T_{\rm perm}$. In this way, $T_{\rm perm}$ represents the level of profile separation expected if the two populations were drawn from the same parent distribution. The significance of the observed separation is then quantified by the probability
\begin{equation}
p = p(T_{\rm perm} > T_{\rm sim}),
\end{equation}
with smaller $p$-values indicating a more significant difference between the `early' and `late' median trends.

For quantities that can take negative values, such as the inflow rate, the separation statistic is evaluated in linear space using $\tilde{y}$ rather than in logarithmic space using $\log_{10}\tilde{y}$. And the same bootstrap and permutation procedure is applied.

\end{document}